\documentclass[prb,twocolumn,superscriptaddress,longbibliography]{revtex4-1}
\RequirePackage[mathlines]{lineno}
\usepackage{amssymb}
\usepackage{amsmath}
\usepackage{mathrsfs}
\usepackage{graphicx}
\usepackage[usenames,dvipsnames]{xcolor}
\definecolor{goodgreen}{rgb}{0.1,0.5,0}
\definecolor{goodred}{rgb}{0.7,0,0}
\usepackage[colorlinks,urlcolor=goodgreen,citecolor=blue,linkcolor=goodred]{hyperref}

\makeatletter
\newsavebox{\@brx}
\newcommand{\llangle}[1][]{\savebox{\@brx}{\(\m@th{#1\langle}\)}%
  \mathopen{\copy\@brx\kern-0.5\wd\@brx\usebox{\@brx}}}
\newcommand{\rrangle}[1][]{\savebox{\@brx}{\(\m@th{#1\rangle}\)}%
  \mathclose{\copy\@brx\kern-0.5\wd\@brx\usebox{\@brx}}}
\makeatother

\usepackage[normalem]{ulem}

\newcommand{\dcom}[1]{\textcolor{NavyBlue}{\textsf{\textbf{#1}}}}

\newcommand{\vect}[1]{\boldsymbol{#1}}

\setpagewiselinenumbers
\modulolinenumbers[5]

\begin{document}

\preprint{\dcom{Preprint not for distribution CONFIDENTIAL, Version of \today}}
\title{Optical activity and transport in twisted bilayer graphene: the essence of spatial dispersion effects}

\author{S. Ta Ho}
\affiliation{Hanoi University of Civil Engineering (HUCE), 55 Giai Phong road, Hanoi 10000, Vietnam}
\author{V. Nam Do}
\email{nam.dovan@phenikaa-uni.edu.vn}    
\affiliation{Department of Basic Science, Phenikaa Institute for Advanced Studies (PIAS), 25$^{th}$ Floor, A9 Building, Phenikaa University, Hanoi 10000, Vietnam}

\begin{abstract}
This study investigates optical activity and quantum transport in twisted bilayer graphene (TBG) systems, demonstrating that the former results from spatial dispersion effects. The transfer matrix method is used to solve the propagation of electromagnetic waves through two graphene layers that act as the coupling surfaces of a dielectric slab. The resulting optical conductivity tensor is decomposed into a local and a drag part, with the drag transverse conductivity $\sigma_{xy}^{(drag)}$ governing the TBG system's optical activity. An effective continuum model is employed to analyze electron state formation and calculate relevant parts of the optical conductivity tensor. Correlation of electron motions leads to incomplete cancellation and a finite $\sigma_{xy}^{(drag)}$ in the chiral TBG lattice. The study also calculates DC conductivity, showing TBG supports quantum conductivity proportional to $e^2/h$ at the intrinsic Fermi energy.
\end{abstract}
\maketitle

\section{Introduction}\label{SecI}
The study of two-dimensional (2D) materials has received significant attention in recent years due to its potential applications in electronics, optoelectronics, spintronics, and valleytronics.\cite{Sierra_2021, Liao_2019, Sangwan_2018, Bao_2019} The use of multiple-layer materials with van der Waals (vdW) interlayer coupling offers various means to tune the electronic structure, such as altering the interlayer distance, changing the lattice alignment through twisting or sliding the layers, and applying external fields. Physically, these material engineering solutions often result in breaking the spatial and temporal symmetries that govern the dynamics of electrons within the material layers. For instance, a magnetic field breaks the time-reversal symmetry, leading to exciting phenomena such as the quantum Hall effect, magneto-optical effects, and the Kerr and Faraday rotations of the polarization plane of linearly polarized light and circular dichroism.\cite{Tse_2011, Crassee_2011, Nandkishore_2011, Shimano_2013} These intriguing optical phenomena have been utilized in the development of various devices for various applications.\cite{Izake_2007, Liu_2005, Solomon_2019} However, the use of external fields has limitations for small-scale device design. As a result, it is crucial to search for materials with intrinsic properties and solutions that break the spatial symmetries of an electronic system to meet technological demands.

Twisted-bilayer graphene (TBG) is a 2D van der Waals (2D vdW) material that has garnered significant attention from the research community due to its interesting electronic properties.\cite{Andrei_2020, Nimbalkar_2020, Hill_2021} The TBG system is formed by stacking two graphene layers on top of each other with a rotation, or twist, angle between the two hexagonal lattices. The TBG system has been shown to be optically active, meaning that it has the ability to rotate the polarization plane of linearly polarized light by reflection and transmission and to absorb left- and right-handed light differently.\cite{Kim_2016} This, combined with the ability to mechanically tune the twist angle,\cite{Palau_2018, Cai_2021} makes TBG a potential candidate for advanced optical applications. The optical activity of a dielectric medium is typically attributed to the anisotropy of its atomic lattice,\cite{Landau_1984} or to the magneto-electric effect.\cite{Condon_1937} External electric, magnetic, and mechanical forces can also induce anisotropy in a medium.\cite{Kim_2016, Nandkishore_2011, Stauber-2018} It has been shown that strained monolayer graphene is optically active due to the presence of the off-diagonal element $\sigma_{xy}(\omega)$ of the optical conductivity tensor.\cite{Olivia_2016} The AA- and AB-stacked configurations of the TBG system are isotropic and characterized by the point groups $D_{6h}$ and $D_{3d}$, respectively.\cite{Zou_2019, Do_2020} The breaking of the lattice symmetry from the $D_{6h}$ and $D_{3d}$ groups to the $D_6$ and $D_3$ groups, respectively, has been suggested as the mechanism causing the optical activity in TBG.\cite{Morell_2017} However, even with this symmetry, the TBG system is characterized by a total diagonal optical conductivity tensor, i.e., with zero off-diagonal components $\sigma_{xy}(\omega)=0$. So, the mechanism supporting the optical activity of TBG remains to be further explored.

In this paper, we analyze the optical activity of twisted bilayer graphene by treating simultaneously two essential aspects of the field-matter interaction problem, namely the propagation of the electromagnetic waves through a material layer and the response of a material layer to the action of an external field. We show that the optical activity of the TBG system is a result of spatial dispersion effects. Our theory is based on the following analysis. Working with 2D van der Waals systems, the physical quantities that describe them, such as the dielectric tensor and the conductivity tensor, are often considered as if they were ideal 2D systems, i.e., with zero thickness, even though the interlayer spacing is taken into account in the description of interlayer coupling.\cite{Moon-2013, Tabert_2013, Stauber_2013} However, in reality, even for graphene with only one layer of carbon atoms, an effective thickness can be defined based on the width of the quantum well potential caused by the carbon atoms.\cite{Do_2017} As a result, 2D vdW systems cannot be considered ideal 2D systems but rather quasi-2D (Q2D) systems. This means that when describing the interaction between an electromagnetic field and a Q2D vdW material system at a macroscopic level, special considerations must be made. Typically, field equations are determined by averaging a set of microscopic field equations over infinitesimal volume elements.\cite{Landau_1984} However, this averaging procedure is only valid when the wavelength of the fields is much smaller than the dimensionalities of the physical system. For Q2D materials, averaging cannot be performed along the $Oz$ direction perpendicular to the $Oxy$ plane of the atomic lattice. As a result, we need to use a hybrid basis of reciprocal-real spaces to represent physical quantities and field equations. For example, the equation
\begin{equation}\label{Eq_1}
    J_\alpha(z,\mathbf{q}_\parallel,\omega) = \int dz^\prime \sigma_{\alpha\beta}(z,z^\prime,\mathbf{q}_\parallel,\omega)E_\beta(z^\prime,\mathbf{q}_\parallel,\omega)
\end{equation}
describes the relationship between the component $J_\alpha(z,\mathbf{q}_\parallel,\omega)$ of an electrical current density induced in the material by an electric field $E_\beta(z,\mathbf{q}_\parallel,\omega)$, where $\alpha, \beta = x,y$ are the indices of the components of vector fields, $\mathbf{q}_\parallel$ is the wave vector in the material plane, $\omega$ is the frequency, and $\sigma_{\alpha\beta}(z,z^\prime,\mathbf{q}_\parallel,\omega)$ is a component of the electrical conductivity tensor of the material layer. In the optical limit, where $\|\mathbf{q}_\parallel\|a\ll 1$ and $a$ is the lattice constant, we can approximate all quantities at $\mathbf{q}_\parallel=0$, i.e., effectively ignoring the spatial effects in the material plane but not along the thickness.\cite{Narazov_2015, Agranovich_2013, Poshakinskiy_2018} This description highlights the importance of taking into account the spatial dispersion along the $Oz$ direction, which is typically the transmission direction of electromagnetic fields used to study the material's properties.

Theoretical studies of optical activity in TBG have been discussed in several references, including \onlinecite{Kim_2016, Morell_2017, Stauber-2018}. These studies have revealed the role of the spacing $a$ between two graphene layers and explained optical activity as a magneto-electric effect. In this approach, a magnetization $\mathbf{m}_\parallel$ and a polarization $\mathbf{p}_\parallel$ emerge in the atomic plane due to the difference in current densities in the two graphene layers. To establish a relationship between these two quantities with electric and magnetic fields, a symmetry analysis and some assumptions were made,\cite{Stauber-2018} such as the linear change of the fields in the space between the two graphene layers and the zero-frequency limit,\cite{Stauber_2018} or the dephasing of the current operators in the two graphene layers.\cite{Morell_2017} The magnetization and polarization affect the direction and magnitude of the electric and magnetic field vectors within the material layer. However, the problem is that the transverse component $\sigma_{xy}^{(drag)}$ of the conductivity tensor, which plays a role in creating the effect of one layer on the other when an electromagnetic field is transmitted through the system, is treated as the total Hall conductivity of the system in this wave propagation problem. This may be misleading as the system does not support a finite Hall conductivity due to time-reversal symmetry. Therefore, it is essential to consider in detail the electromagnetic wave propagation through the system, where it is understood that the electromagnetic wave must pass through one layer before reaching the second layer, and the coupling between the two layers must play a crucial role. In the next section, we will use Eq. (\ref{Eq_1}) without considering symmetry analysis and the mentioned assumptions to obtain the most important theoretical results in Ref. (\onlinecite{Stauber-2018}), see Eq. (2) therein, and clarify the role of $\sigma_{xy}^{(drag)}$ in the electromagnetic wave propagation through the TBG system.

Our study encompasses both macroscopic and microscopic perspectives. At the macroscopic level, we use the transfer matrix method to analyze the light propagation through a 2D van der Waals system composed of multiple material layers, each treated as a boundary between two vacuum layers of finite width.\cite{Zhan_2013,Szechenyi_2016} The material system's conductivity tensor is decomposed into local and drag terms, with the local term defining the local relationship between current density and electric field in each layer, and the drag term defining the non-local relationship between current density in one layer and the electric field in other layers. Technically, the finite thickness of the material system makes the assumption of the spatial homogeneity along the thickness invalid. It therefore requires evaluating expressions that relate physical quantities through integrals over $z$ and $z'$. This results in increased computational complexity and the need for a microscopic description of the material surfaces, such as the potential caused by atoms. To simplify the analysis, we assume that each layer of atoms has zero thickness. It allow us to express relevant quantities as sums of delta functions:
\begin{equation}\label{Eq_2}
J_\alpha(z,\omega) = \sum_\ell J_\alpha^{(\ell)}\delta(z-z_\ell),
\end{equation}
where $z_\ell$ is the position of the $\ell$th material layer, and $J_\alpha^{(\ell)}$ is the current density on that layer.

At the microscopic level, we use the Kubo formula to calculate each term of the conductivity tensor required for the macroscopic level of description. We apply an effective model to determine the low-energy states of electrons in the TBG material lattice, and show that the electron eigen-states can be expressed as a hybridization of single-layer states in the two graphene layers. If the atomic lattice has mirror symmetry (the AA-stacked configuration) or glide symmetry (the AB-stacked configuration) in the lattice plane, the drag transverse conductivity will be vanished due to the cancelling contribution of the electron states. However, generic twisted bilayer graphene lattices have a chiral structure and do not possess these symmetries, leading to a non-zero drag transverse conductivity that determines the optical activity of the system. In addition to determining the components of the optical conductivity tensor, the reflection, transmission, absorption coefficients, and the circular dichroism spectrum, we also calculate the DC conductivity of the TBG system. We find that the hybridized states support a quantum value of $2\sigma_0^{DC}$, where $\sigma_0^{DC} = \frac{4e^2}{\pi h}$, at the intrinsic Fermi energy. This holds true even for the TBG configuration with a magic twist angle, which has a high density of states there due to the presence of a flat band.

The structure of this paper is divided into three sections. After the ``Introduction" section, the next section (Sec. \ref{SecII}) provides a solution to the problem of wave transmission through the twisted bilayer graphene (TBG) system in Sec. \ref{SecIIA}. The solution is based on the transfer matrix method, which illustrates the contribution of the conductivity tensor components to the transfer matrix elements such as the reflection and transmission coefficients and the absorption coefficient. In Sec. \ref{SecIIB}, the local and drag components of the optical conductivity tensor are calculated through the use of the Kubo formula, while the DC conductivity of the TBG system is calculated using the Kubo-Greenwood formula. In Sec. \ref{SecIIC}, an effective model for the electrons in the TBG lattices is presented. In Sec. \ref{SecIII}, numerical results for the electronic band structure of various TBG configurations, the analysis of the optical conductivity tensor elements, and the spectra of circular dichroism as a function of photon frequency are presented. The conclusion is presented in the final section, Sec. \ref{SecIV}.
\begin{figure*}\centering
\includegraphics[clip=true,trim=1.7cm 7cm 7cm 7cm,width=0.4\textwidth]{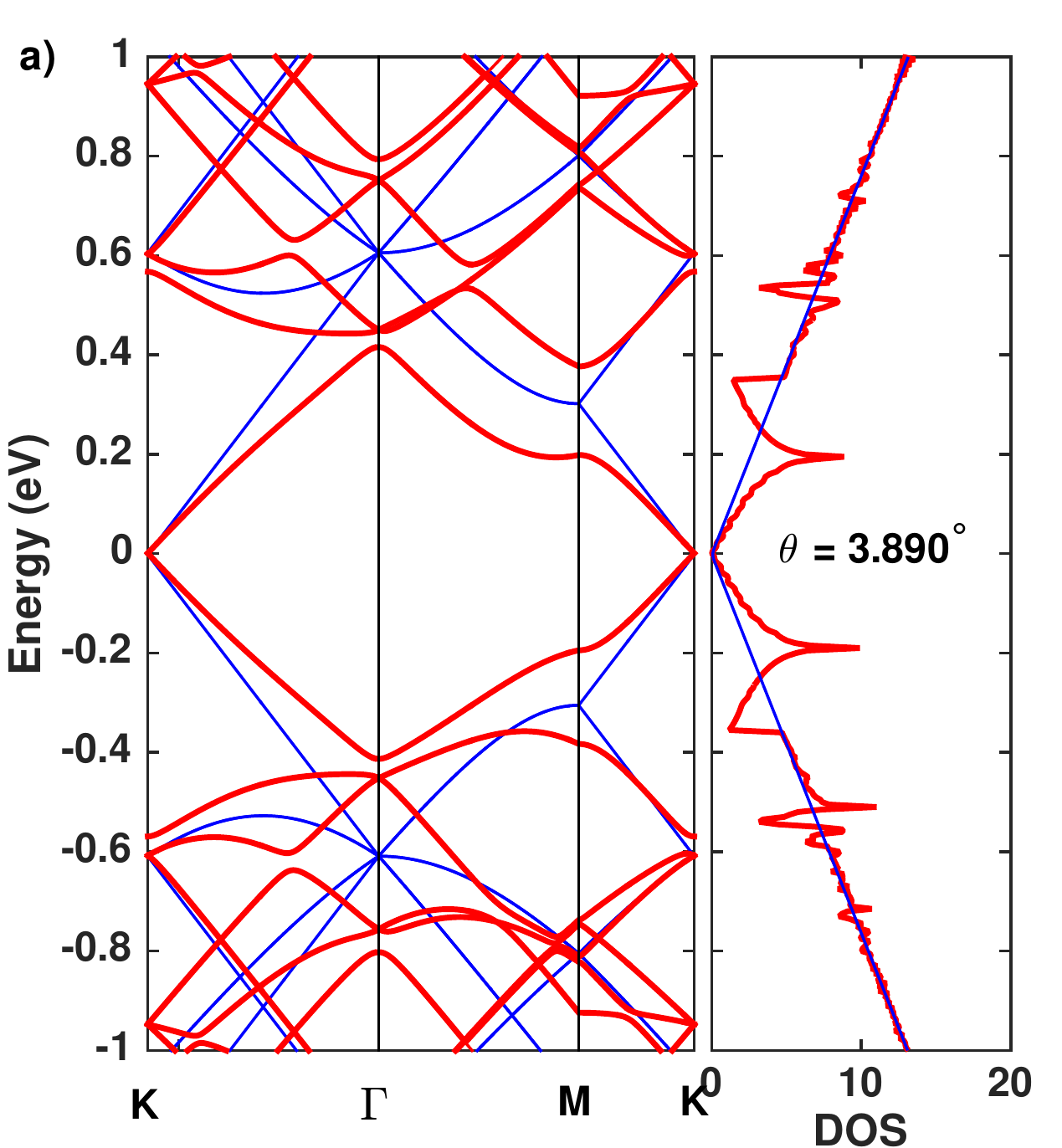}
\includegraphics[clip=true,trim=1.7cm 7cm 11.36cm 7cm,width=0.265\textwidth]{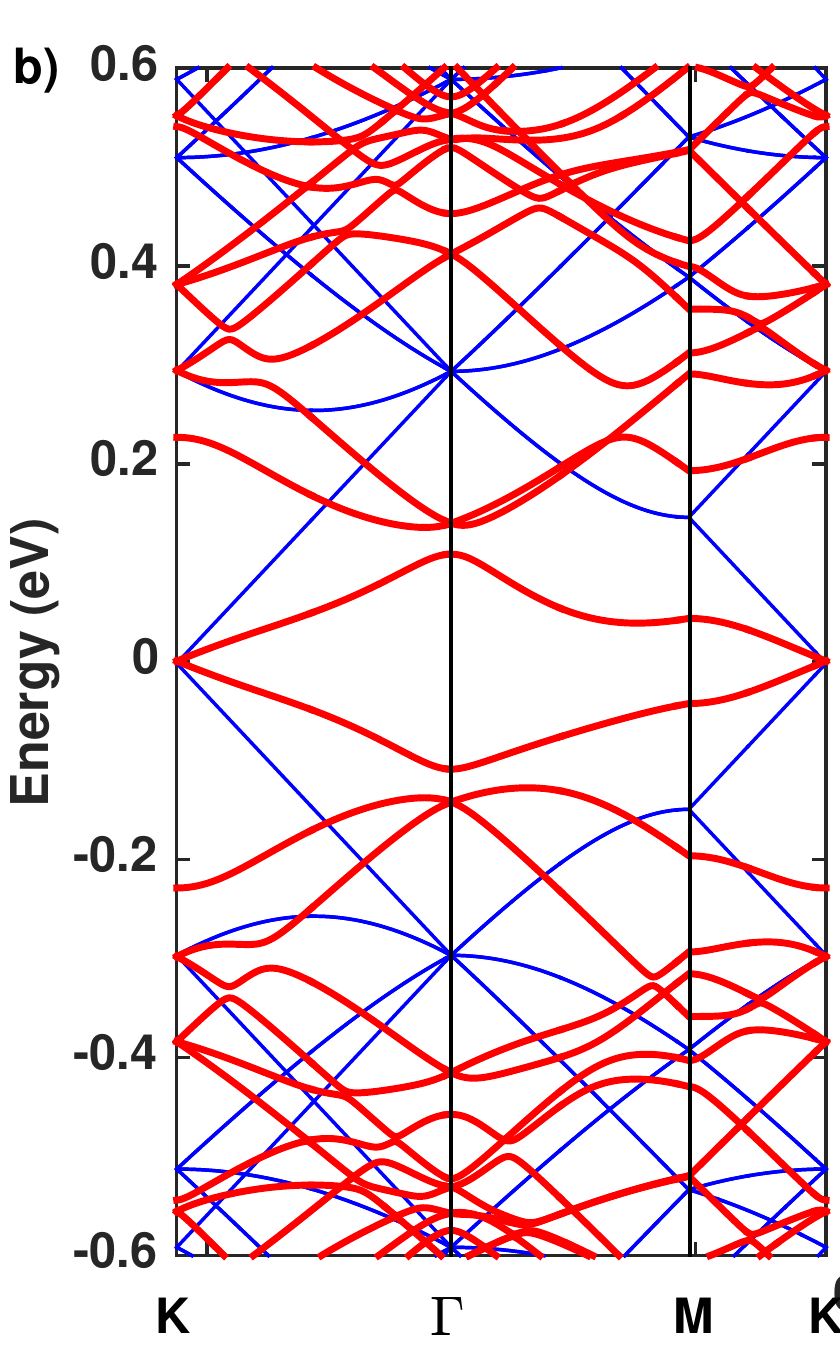}\\
\includegraphics[clip=true,trim=1.7cm 7cm 7cm 7cm,width=0.4\textwidth]{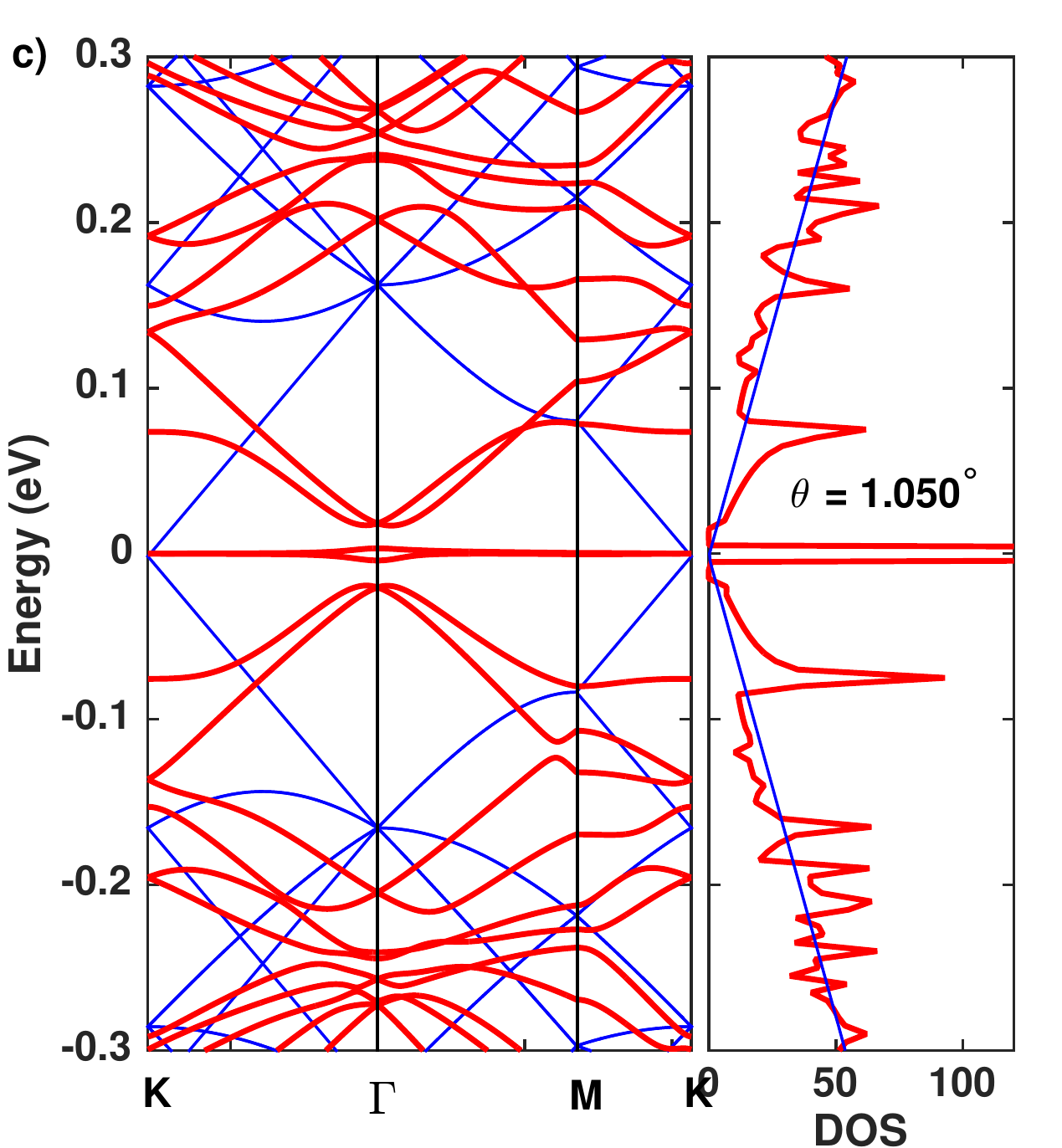}
\includegraphics[clip=true,trim=1.7cm 7cm 11.36cm 7cm,width=0.265\textwidth]{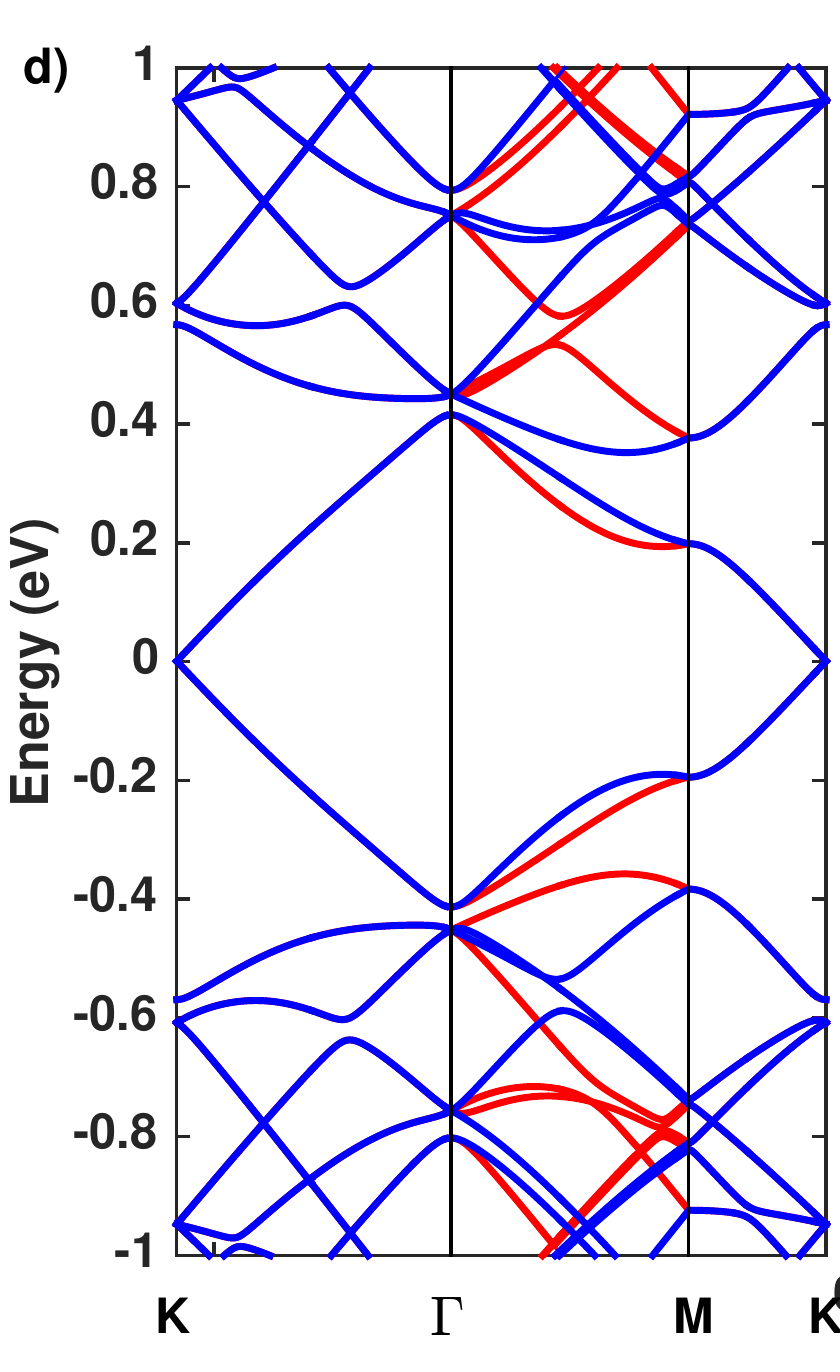}\\
\includegraphics[clip=true,trim=1.7cm 7cm 7cm 7cm,width=0.4\textwidth]{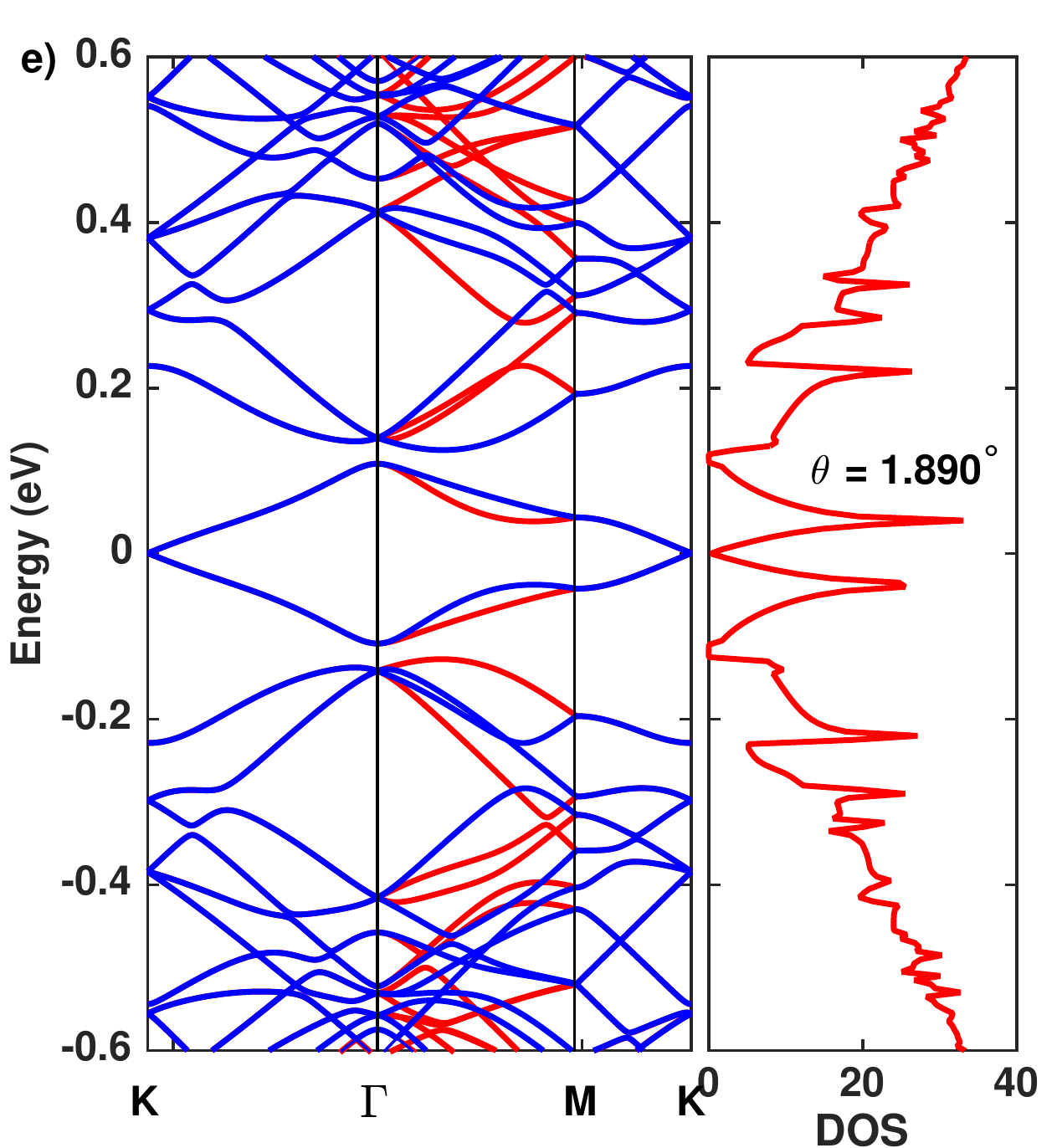}
\includegraphics[clip=true,trim=1.7cm 7cm 11.36cm 7cm,width=0.265\textwidth]{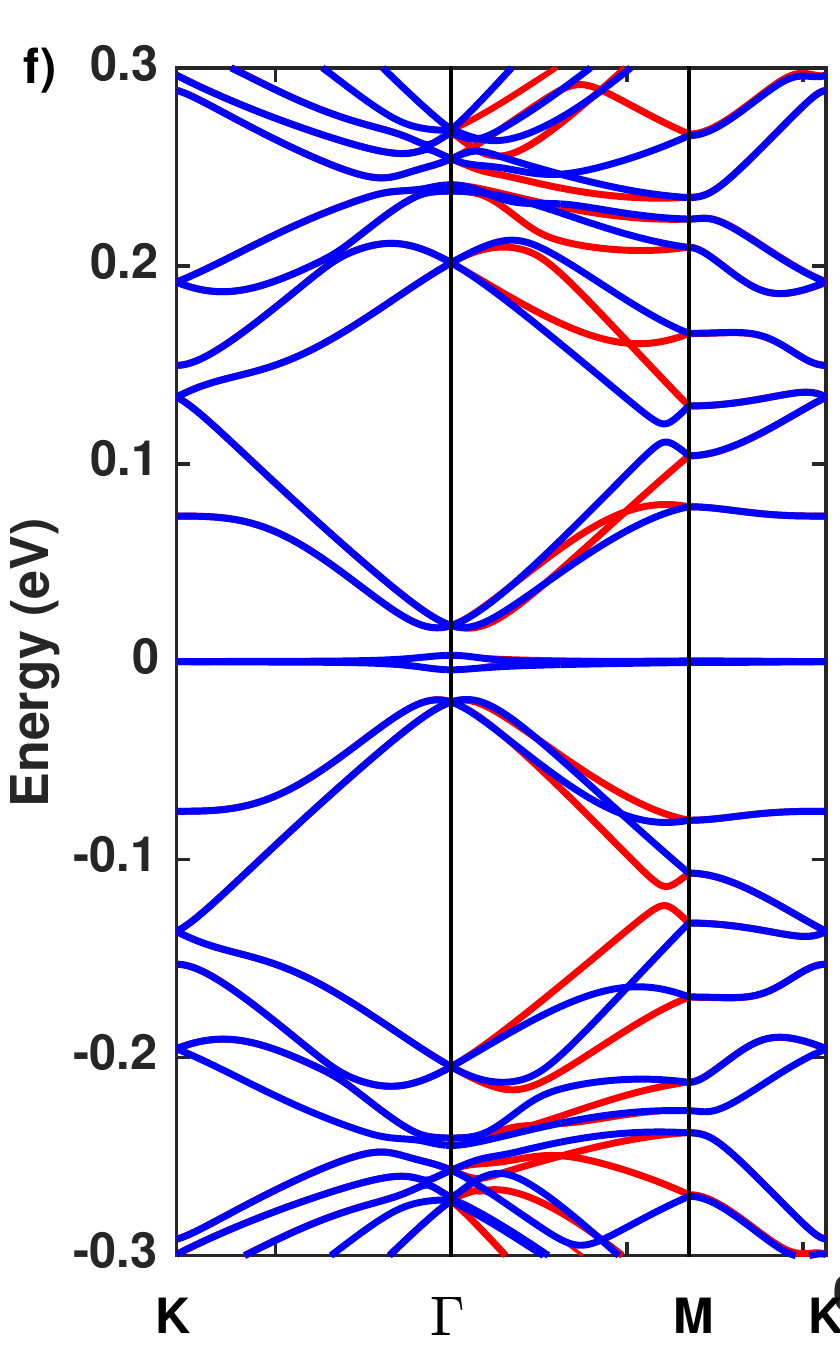}
\caption{\label{Fig1} The energy band structures of electrons in three TBG configurations with twist angles $\theta = 3.890^\circ$ (a, d), $\theta = 1.890^\circ$ (b, e), and $\theta = 1.050^\circ$ (c, f) are analyzed. The blue curves in (a), (b), and (c) represent the dispersion and density of states of graphene calculated without considering the interlayer coupling. Panels (d), (e), and (f) show the dispersion curves for the two valleys $\xi = +1$ and $\xi = -1$ in blue and red, respectively.}
\end{figure*}
\section{Theory and calculation methods}\label{SecII}
\subsection{Optical chiral responses}\label{SecIIA}
In this section, we investigate the optical response of TBG systems by formulating a theory for the propagation of electromagnetic waves through a generic TBG sheet. Unlike typical dielectric slabs with two separate surfaces, the electric current densities in the two graphene layers of the TBG sheet are interdependent. The current density in one graphene layer is affected by both the electric field within that layer and the electric field in the other layer. This relationship is described by Eq. (\ref{Eq_1}), which we use to solve the problem at hand.

We set up the system, following Ref.~\onlinecite{Szechenyi_2016}, with two graphene layers separating the entire space into three regions, each characterized by parameters $(\epsilon_i,\mu_i)$, where $i=1,2,3$ labels the mediums. Our goal is to determine the transfer matrix of the system and to develop expressions for the transmission and reflection matrices of the bilayer graphene system. As the inhomogeneity of the space only occurs in the $z$ direction perpendicular to the TBG surface, the Maxwell equations support a plane wave solution of the form:
\begin{equation}\label{Eq25}
\vect{F}_i(x,y,z) = \left[\vect{F}_i^+e^{ik_{iz}z} + \vect{F}_i^{-}e^{-ik_{iz}z}\right]e^{ik_{ix}x}e^{ik_{iy}y},
\end{equation}
where $\vect{F}$ represents $\vect{E}$ (the electric field), $\vect{D}$ (the electric induction), $\vect{H}$ (the magnetic field), and $\vect{B}$ (the magnetic induction). The vector coefficients $\vect{F}_i^\pm$ depend on the wave vector $\vect{k}_i$ in general. The conditions for the continuity of the vector fields at the interface between two mediums are given by:
\begin{subequations}
\begin{align}
&\vect{n}\times[\vect{E}_2(1)-\vect{E}_1(1)] = 0,\label{Eq26a}\\
&\vect{n}\times[\vect{H}_2(1)-\vect{H}_1(1)] = \vect{J}(1),\label{Eq26b}
\end{align}
\end{subequations}
for the first graphene layer surface, and
\begin{subequations}
\begin{align}
&\vect{n}\times[\vect{E}_3(2)-\vect{E}_2(2)] = 0,\label{Eq27a}\\
&\vect{n}\times[\vect{H}_3(2)-\vect{H}_2(2)] = \vect{J}(2).\label{Eq27b}
\end{align}
\end{subequations}
for the second graphene layer surface. In Eqs. (\ref{Eq26b}) and (\ref{Eq27b}), the distribution of current densities on each graphene layer is linearly related to the electric field by the conductivity tensor given via Eq. (\ref{Eq5}). To the completeness of the setup, we rewrite this equation in the explicit form:
\begin{subequations}
\begin{align}
    \vect{J}(1) &= \vect{\sigma}^{(1)}\vect{E}_1(1)+\vect{\sigma}^{(drag)}\vect{E}_2(2),\label{Eq28a}\\
    \vect{J}(2) &= \vect{\sigma}^{(drag)\dagger}\vect{E}_1(1)+\vect{\sigma}^{(2)}\vect{E}_2(2)\nonumber\\
    &=\vect{\sigma}^{(drag)\dagger}\vect{E}_2(1)+\vect{\sigma}^{(2)}\vect{E}_3(2).\label{Eq28b}
\end{align}
\end{subequations}

The continuity conditions of the fields at interfaces allow us to determine the relationship between the vector fields $(\vect{F}_i^+,\vect{F}_i^-)$ in two different media at their interface. In particular, the magnetic field $\vect{H}_i$ is related to the electric field $\vect{E}_i$ through the Maxwell equation $\text{curl}(\vect{E}) = -\partial_t\vect{B}$, where $\vect{B}=\mu\vect{H}$. Utilizing the plane wave solution (\ref{Eq25}), we obtain the expression for $\vect{H}_i^\pm$:
\begin{equation}\label{Eq29}
\vect{H}_i^\pm = \pm\sqrt{\frac{\epsilon_i}{\mu_i}}\frac{\vect{k}_i\times\vect{E}_i^\pm}{|\vect{k}_i|},
\end{equation}
where $\epsilon_i$ and $\mu_i$ represent the absolute permittivity and absolute permeability of medium $i$, respectively. This equation makes use of the dispersion relation $\omega = c|\vect{k}_i|/n_i$, where $c$ is the speed of light in vacuum and $n_i$ is the refractive index of medium $i$. It is worth noting that $\sqrt{\epsilon_i/\mu_i}$ can be expressed through $n_i$ as $\sqrt{\epsilon_i/\mu_i} \approx n_i\sigma_0/2\alpha$, where $\sigma_0=e^2/h$ is the unit of quantum conductivity and $\alpha = e^2/(4\pi\epsilon_0\hbar c)$ is the fine-structure constant. In this study, the problem is solved for light transmission along the normal direction to the TBG layer, so $\vect{k}_i = k_i\vect{n}$. The transfer matrix $\mathbf{M}{31}$ (of size $4\times 4$) is then derived, relating the components of the electric fields in the first and third mediums as follows:
\begin{align}\label{Eq30}
\left(\begin{array}{c}
\vect{E}_3^+(2) \\
\vect{E}3^-(2)
\end{array}\right)
&= \mathbf{M}{31} \left(\begin{array}{c}
\vect{E}_1^+(1) \\
\vect{E}_1^-(1)
\end{array}\right).
\end{align}
Here, $\mathbf{M}_{31}=\mathbf{M}_{32}\mathbf{M}^f_2\mathbf{M}_{21}$, where:
\begin{widetext}
\begin{subequations}
\begin{align}
    \mathbf{M}_{32} &= \left(\begin{array}{cc}
        \vect{1} & \vect{1} \\
        -\sqrt{\frac{\epsilon_3}{\mu_3}}\vect{1}-\vect{\sigma}^{(2)} &\sqrt{\frac{\epsilon_3}{\mu_3}}\vect{1}-\vect{\sigma}^{(2)} 
    \end{array}\right)^{-1}\left(\begin{array}{cc}
        \vect{1} & \vect{1} \\
        -\sqrt{\frac{\epsilon_2}{\mu_2}}\vect{1}+\vect{\sigma}^{(drag)\dagger}e^{-ik_{2z}d} & \sqrt{\frac{\epsilon_2}{\mu_2}}\vect{1}+\vect{\sigma}^{(drag)\dagger}e^{ik_{2z}d} 
    \end{array}\right), \label{Eq31a}\\
        \mathbf{M}_{21} &= \left(\begin{array}{cc}
    \vect{1}&\vect{1}\\
    -\sqrt{\frac{\epsilon_2}{\mu_2}}\vect{1}-\vect{\sigma}^{(drag)}e^{ik_{2z}d}&\sqrt{\frac{\epsilon_2}{\mu_2}}\vect{1}-\vect{\sigma}^{(drag)}e^{-ik_{2z}d}\end{array}\right)^{-1}\left(\begin{array}{cc}
      \vect{1}   & \vect{1} \\
    -\sqrt{\frac{\epsilon_1}{\mu_1}}\vect{1}+\vect{\sigma}^{(1)}     & \sqrt{\frac{\epsilon_1}{\mu_1}}\vect{1}+\vect{\sigma}^{(1)}
    \end{array}\right), \label{Eq31b}\\
    \mathbf{M}_2^f &=\left(\begin{array}{cc}
    e^{ik_{2z}d}\vect{1} & 0 \\
        0 & e^{-ik_{2z}d}\vect{1} 
    \end{array}\right), \label{Eq31c}
\end{align}
\end{subequations}
\end{widetext}
with $\vect{1}$ the $2\times 2$ identity matrix.

Let $\mathbf{r}$ and $\mathbf{t}$ be the $2\times 2$ matrices that relate the components of the electric field vector of the incident light to those of the reflected and transmitted light, i.e., $\vect{E}_1^-(1) = \mathbf{r}\cdot\vect{E}_1^+$ and $\vect{E}_3^+=\mathbf{t}\cdot\vect{E}_1^+$. These matrices are referred to as the reflection and transmission matrices, respectively. From Eq. (\ref{Eq30}), we can deduce the following:
\begin{subequations}
\begin{align}
\mathbf{r} &= -[\mathbf{M}_{31}^{22}]^{-1}\mathbf{M}_{31}^{21}, \label{Eq32a}\\
\mathbf{t} &= \mathbf{M}_{31}^{11}-\mathbf{M}_{31}^{12}[\mathbf{M}_{31}^{22}]^{-1}\mathbf{M}_{31}^{21}=[\mathbf{M}_{31}^{-1}]^{11}. \label{Eq32b}
\end{align}
\end{subequations}
From these matrices, the reflectance ($R$) and transmittance ($T$) are determined as the ratio of energy fluxes, specifically:
\begin{subequations}
\begin{align}
R &= \left\vert\frac{(\mathbf{r}\cdot\vect{E}_1^+)^\dagger\cdot(\mathbf{r}\cdot\vect{E}_1^+)}{\vect{E}_1^{+\dagger}\cdot\vect{E}_1^+}\right\vert, \label{Eq33a}\\
T &= \left\vert\frac{n_3}{n_1}\frac{(\mathbf{t}\cdot\vect{E}_1^{+})^\dagger\cdot(\mathbf{t}\cdot\vect{E}_1^+)}{\vect{E}_1^{+\dagger}\cdot\vect{E}_1^+}\right\vert. \label{Eq33b}
\end{align}
\end{subequations}
On the base of the conservation of energy flux, the absorptance is then determined by $A = 1-(R+T)$. The values of $R$, $T$, and $A$ are dependent not only on the energy, but also on the polarization states of light. It is important to note that the left- and right-handed polarization states of light are characterized by electric field vectors, $\vect{E}_{L,R}^+ \propto (1,\pm i)^T/\sqrt{2}$ (The supersript $T$ here denotes the transpose operation). To measure the dependence of light absorption on these polarization states, we calculate the spectrum of circular dichroism (CD), defined as:
\begin{equation} \label{Eq34}
\mathrm{CD} = \frac{A_\mathrm{L}-A_\mathrm{R}}{A_\mathrm{L}+A_\mathrm{R}}.
\end{equation}
In the following section, we present the method for calculating the conductivity tensors $\boldsymbol{\sigma}^{(\ell)}$ and $\boldsymbol{\sigma}^{(drag)}$, and we will use the results obtained to calculate the transmittance, reflectance, absorbance and circular dichroism via Eqs. (\ref{Eq33a}), (\ref{Eq33b}) and (\ref{Eq34}).
\begin{figure}\centering
\includegraphics[clip=true,trim=1.5cm 6.5cm 2cm 7cm,width=\columnwidth]{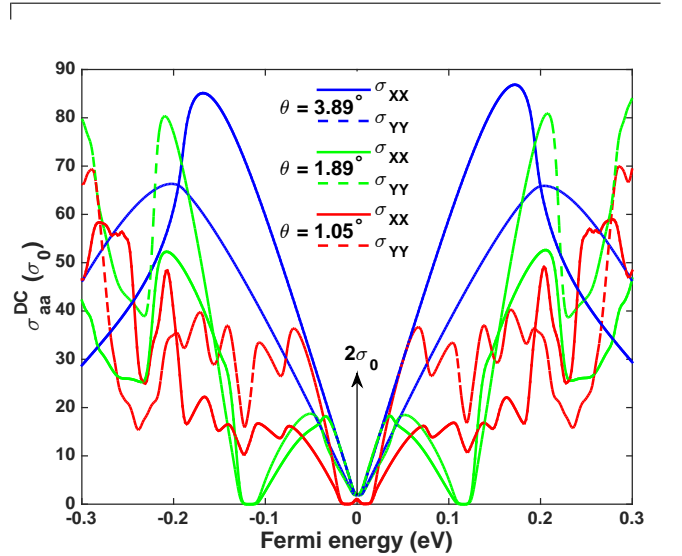}
\caption{\label{Fig2} The longitudinal DC conductivities $\sigma_{xx}^{DC}$ (represented by solid curves) and $\sigma_{yy}^{DC}$ (represented by dashed curves) for three TBG configurations with twist angles of $\theta = 3.890^\circ$ (represented by blue curves), $\theta = 1.890^\circ$ (represented by green curves), and $\theta = 1.050^\circ$ (represented by red curves).}
\end{figure}
\subsection{AC and DC conductivities}\label{SecIIB}
The optical properties of bilayer graphene systems are notable, as evidenced by previous research.\cite{Moon-2013, Stauber-2018, Do_2018, Catarina_2019} When taking into account the layer thickness, the effects of spatial dispersion become evident. To describe these effects, we decompose the current response to an external electric field caused by an incident monochromatic light beam perpendicular to the material plane. To start, we rewrite Equation (\ref{Eq_1}) in the real space representation as follows:

\begin{align}\label{Eq1}
J_\alpha(\vect{x},z, \omega) &= \int dz^\prime\sigma_{\alpha\beta}(\vect{x},z;\vect{x},z^\prime,\omega)E_\beta(\vect{x},z^\prime,\omega),
\end{align}
where $\mathbf{x} = (x,y)$ represents the position vector in the material plane with $x$ and $y$ as its coordinates. On the assumption of spatial homogeneity in the material plane, we can write $\sigma_{\alpha\beta}(\vect{x},z;\vect{x},z^\prime, \omega) = \sigma_{\alpha\beta}(z,z^\prime,\omega)$, and:
\begin{align}\label{Eq2}
    \sigma_{\alpha\beta}(z,z^\prime,\omega)  =&\sum_{\ell=1}^2\sigma_{\alpha\beta}(z,z^\prime,\omega)\delta(z^\prime-z_\ell).
\end{align}
Eq. (\ref{Eq1}) therefore becomes:
\begin{align}\label{Eq3}
    J_\alpha(\vect{x},z,\omega) = \sum_{\ell=1}^2\sigma_{\alpha\beta}(z,z_\ell,\omega)E_\beta(\vect{x},z_\ell,\omega).
\end{align}
More specifically, we resolve the current density induced in each graphene layer as follows:
\begin{equation}\label{Eq4}
    J_\alpha^{(\ell)}(\vect{x},\omega) = \sigma_{\alpha\beta}^{(\ell)}(\omega)E_{\beta}^{(\ell)}(\vect{x},\omega)+\sigma_{\alpha\beta}^{(drag)}(\omega)E_\beta^{(\ell^\prime)}(\vect{x},\omega),
\end{equation}
where the involved quantities are defined by: $J_\alpha^{(\ell)}(\vect{x},\omega) = J_\alpha(\vect{x},z_\ell,\omega); E_\beta^{(\ell)}(\vect{x},\omega) = E_\beta(\vect{x},z_\ell,\omega)$ and $\sigma_{\alpha\beta}^{(\ell)}(\omega) = \sigma_{\alpha\beta}(z_\ell,z_\ell,\omega)$, and  $\sigma_{\alpha\beta}^{(drag)}(\omega) = \sigma_{\alpha\beta}(z_\ell,z_{\ell^\prime},\omega)$ where $\ell\neq\ell^\prime$. Eqs. (\ref{Eq4}) can be cast in the matrix form:
\begin{align}\label{Eq5}
    \left(\begin{array}{c}
    \vect{J}^{(1)}\\ \vect{J}^{(2)}
    \end{array}\right)_{\omega} = \left(\begin{array}{cc}
    \vect{\sigma}^{(1)} & \vect{\sigma}^{(drag)}\\
    \vect{\sigma}^{(drag)\dagger} &\vect{\sigma}^{(2)}
    \end{array}\right)_\omega\left(\begin{array}{c}
    \vect{E}^{(1)}\\ \vect{E}^{(2)}
    \end{array}\right)_{\omega},
\end{align}
where $\vect{\sigma}^{(\ell)}(\omega) = \sigma^{(\ell)}(\omega)\vect{\tau}_0$ is called the local part, and  $\vect{\sigma}^{(drag)}(\omega) = \sigma^{(drag)}(\omega)\vect{\tau}_0-i\sigma_{xy}^{(drag)}(\omega)\vect{\tau}_y = \sigma^{drag}(\omega)-\sigma_{xy}^{drag}\hat{\vect{z}}\times$ is called the drag part. Here $\vect{\tau}_0$ is the $2\times 2$ identity matrix and $\vect{\tau}_2$ is the conventional second Pauli matrix. The compact form (\ref{Eq5}) is identical to results reported in Refs. \onlinecite{Stauber-2018,Ochoa_2020}. Importantly, it guarantees the time-reversal symmetry, the rotation symmetry and the layer interchange symmetry. 

The conductivities $\sigma^{(\ell)}(\omega), \sigma^{(drag)}(\omega)$ and $\sigma^{(drag)}_{xy}(\omega)$ can be determined from the Kubo formula. There are a number of versions of the Kubo formula for the electrical conductivity suitable for implementing it in different situations. However, the most important ingredient we must specify is the velocity operators $\hat{v}_\alpha$. In general, these operators are determined from the position operator $\hat{x}_\alpha$ and the Hamiltonian $\hat{H}$ via the Heisenberg equation:
\begin{equation}\label{Eq6}
    \hat{v}_\alpha = \frac{1}{i\hbar}[\hat{x}_\alpha,\hat{H}] \rightarrow v_\alpha(\vect{k})= \frac{1}{\hbar}\frac{\partial}{\partial k_\alpha}H(\vect{k}).
\end{equation}
According to the linear response theory, we see that the conductivity has the paramagnetic part only. Using the eigen-vectors of single-particle Hamiltonian (without the presence of external vector potential $\vect{A}(\vect{x},t)$) as the representation basis, $\hat{H}|n\rangle = E_n|n\rangle$, the elements of the optical conductivity tensor are given by this formula:~\cite{Allen_2006}
\begin{align}\label{Eq7}
\sigma_{\alpha\beta}^{(c)}(\omega) &= \frac{i  e^2}{S}\frac{1}{\omega+i\eta}\sum_{m,n}(f_m-f_n)\frac{O^{(c)mn}_{\alpha\beta}}{E_m-E_n+\hbar(\omega+i\eta)},
\end{align}
where $S$ is the area of material sample; $c = \{\ell,drag\}$; $f_n = f(E-\mu,k_BT)$ is the occupation weight of the energy level $E_n$ determined by the Fermi-Dirac function $f$ with $\mu,k_BT$ the chemical potential and thermal energy; $\eta$ is a positive infinitesimal number, and $O_{\alpha\beta;mn}^{(c)}$ denotes the product of velocity matrix elements:
\begin{subequations}
\begin{align}
    O_{\alpha\beta}^{(\ell)mn} &= \langle m|\hat{v}_\alpha^{(\ell)}|n\rangle\langle n|\hat{v}_\beta^{(\ell)}|m\rangle,\label{Eq8a} \\
    O_{\alpha\beta}^{(drag)mn} &= \langle m|\hat{v}_\alpha^{(1)}|n\rangle\langle n|\hat{v}_\beta^{(2)}|m\rangle\label{Eq8b}.
\end{align}
\end{subequations}

To the DC conductivity, it might be optimistic to think that the DC conductivity can be simply obtained from the expression of the optical conductivity in the limit of zero frequency. However, it is not, at least in the numerical calculation. This is because for some finite frequency $\omega$, there is always a finite length scale governing the behavior of electron. This length scale is $L_\omega = 2\pi v_F/\omega$. Meanwhile, there is no such length scale for the DC transport. Furthermore, the static transport has the diffusion nature due to the similarity of the Schr\"odinger equation and the diffusion equation. The DC conductivity can be obtained from the linear response theory. The vector potential is chosen in the form $\vect{A}(t) = (-Et,0,0)$, where $E$ is the intensity of a static electric field. The DC conductivity is calculated using the Kubo-Greenwood formula:\cite{Kubo_1957,Greenwood_1958}
\begin{equation}\label{Eq9}
    \sigma_{\alpha\alpha}(\mu,kT) = -\int_{-\infty}^{+\infty}dE \frac{\partial f(E-\mu,kT)}{\partial E}\sigma_{\alpha\alpha}(E),
\end{equation}
where
\begin{equation}\label{Eq10}
    \sigma_{\alpha\alpha}(E) = \frac{2\pi e^2\hbar}{S}\sum_{m,n}|\langle m|\hat{v}_\alpha|n\rangle|^2 \delta(E-E_n)\delta(E-E_m).
\end{equation}
This quantity is seen as the DC conductivity at zero temperature. When performing the above formula, we use the Gaussian function to approximate the $\delta$-Dirac function:
\begin{equation}\label{Eq11}
    \delta(E-E_n) \approx \frac{1}{\eta\sqrt{\pi}}\exp\left(-\frac{(E-E_n)^2}{\eta^2}\right),
\end{equation}
where $\eta > 0$ is a tiny number that is appropriately chosen to smear the energy levels for the numerical calculation.
\subsection{Effective continuum model}\label{SecIIC}
Effective continuum models for the low-energy states of electrons in TBG systems have been developed since 2007 by Lopes et al. \cite{Santos_2007} However, the model proposed by Bistritzer and MacDonald in 2011 is better known and widely used. In this paper, we present our solution to the Bistritzer-MacDonald model. \cite{Bistritzer_2011} We implement this model for TBG lattices in which the first layer is clockwise rotated by a half-twist angle, $\theta_1 = -\theta/2$, and the second layer is counterclockwise rotated by $\theta_2 = +\theta/2$. The unrotated layers are defined by unit vectors $\vect{a}_1 = a\cdot\vect{e}_x$ and $\vect{a}_2 = a\cos\pi/3\cdot\vect{e}_x + a\sin\pi/3\cdot\vect{e}_y$, where $a$ is the lattice constant of graphene, and $\vect{e}_x$ and $\vect{e}_y$ are the unit vectors defining the $x$ and $y$ directions, respectively. The two associated vectors defining the reciprocal lattice are thus $\vect{a}_1^\star = (4\pi/a)(\cos\pi/6\cdot\vect{e}_x+\sin\pi/6\cdot\vect{e}_y)$ and $\vect{a}_2^\star = (4\pi/a)\cdot\vect{e}_y$. Under the twisting, these vectors defining the atomic lattice of the two graphene layers as well as their reciprocal lattice are determined by applying the rotation matrix $R_z(\theta_\ell)$, specifically $\vect{a}_i^{(\ell)} = R_z(\theta_\ell)\cdot\vect{a}_i$, and $\vect{a}_i^{\star(\ell)} = R_z(\theta_\ell)\cdot\vect{a}_i^{\star}$. The first Brillouin zone $\text{BZ}^{(\ell)}$ of each graphene layer is defined by a set of six corner points $\vect{K}_i^{(\ell)}$. For instance, the point $\vect{K}_{1}^{(\ell)}$ is determined by the point $\vect{K}_{-}^{(\ell)}$ with $\vect{K}_{\xi}^{(\ell)} = -\xi(2\vect{a}_1^{\star(\ell)}+\vect{a}_2^{\star(\ell)})/3$, where $\xi = \pm 1$.

For commensurate twisting, the twist angle $\theta$ is determined by two integer numbers $m$ and $n$ via the formula:
\begin{equation}\label{Eq12}
\theta = \arctan\left(\frac{|n^2-m^2|\sin(\pi/3)}{(n^2+m^2)\cos(\pi/3)+2mn}\right).
\end{equation}
When the absolute value of the difference between $m$ and $n$ is equal to 1, two unit vectors $\vect{A}_i^\star = \vect{a}_i^{\star(1)}-\vect{a}_i^{\star(2)}$ define the moiré reciprocal lattice. The mini-Brillouin zone comprises six $\vect{K}^M_i$ points. $\vect{K}_1^M = (-\vect{A}_1^\star+\vect{A}_2^\star)/3$, and $\vect{K}^M_6 = (\vect{A}_1^\star+2\vect{A}_2^\star)/3$ define two of these points. The remaining points can be obtained by shifting these points using the reciprocal unit vectors.

We begin by representing the Hamiltonian $H$ for electrons using the layer-resolution vector basis set $\{|\ell\rangle,|,\ell = 1,2\}$. Using the identity $1=\sum_{\ell=1}^2|\ell\rangle\langle\ell|$, we can express $H$ as follows:
\begin{equation}\label{Eq13}
H = \sum_{\ell,\ell^\prime=1}^2|\ell\rangle H_{\ell,\ell^\prime}\langle\ell^\prime|,
\end{equation}
where $H_{\ell,\ell^\prime} = \langle\ell|H|\ell^\prime\rangle$. Next, we can further specify $H_{\ell,\ell^\prime}$ by using the lattice-resolution vector basis set ${|\vect{k},m\rangle = |\vect{k}+\vect{G}_m\rangle ,|,\vect{k}\in BZ, m\in \mathbb{Z}}$. We can then use the identity $1=\sum{\vect{k}\in BZ}\sum_m|\vect{k},m\rangle\langle\vect{k},m|$ to obtain:
\begin{align}\label{Eq14}
H
&= \sum_{\vect{k}\in BZ}\sum_{\ell,m}\sum_{\ell^\prime,n}|\ell,\vect{k},m\rangle H_{\ell,m;\ell^\prime,n}(\vect{k})\langle\ell^\prime,\vect{k},n|,
\end{align}
where $|\ell,\vect{k},m\rangle = |\ell\rangle|\vect{k},m\rangle$ and we use the relation $\langle\ell,\vect{k},m|H|\ell^\prime,\vect{k}^\prime,n\rangle = H_{\ell,m;\ell^\prime,n}(\vect{k})\delta_{\vect{k},\vect{k}^\prime}$.
In the case of twisted bilayer graphene (TBG) configurations with tiny twist angles, the low energy states of electrons are distinguished by a quantum index $\xi = \pm1$, which corresponds to the two nonequivalent Dirac valleys $\vect{K}_{\xi}^{(1,2)}$ of the graphene monolayers. As the twist angle becomes smaller, the positions of the two points $\vect{K}_{\xi}^{(1)}$ and $\vect{K}_{\xi}^{(2)}$ become closer to each other. The Bistritzer-MacDonald model is represented by a Hamiltonian in real-space representation as follows:\cite{Bistritzer_2011}
\begin{align}\label{Eq15}
\hat{H}^\xi =
\begin{pmatrix}
H_1^\xi(\hat{\vect{p}}) & T^\xi(\hat{\vect{r}}) \\
T^{\xi\dagger}(\hat{\vect{r}}) & H_2^\xi(\hat{\vect{p}})
\end{pmatrix}.
\end{align} 
Here, $H_1^\xi(\hat{\vect{p}})$ and $H_2^\xi(\hat{\vect{p}})$ are the blocks for the Hamiltonians of the two monolayers, and $T^\xi(\hat{\vect{r}})$ is the interlayer coupling block, which are determined via the momentum and position operators, $\hat{\vect{p}}$ and $\hat{\vect{r}}$, respectively. For the range of energy around the intrinsic Fermi level the two Hamiltonian blocks $H_1^\xi(\hat{\vect{p}})$ and $H_2^\xi(\hat{\vect{p}})$  are well approximated by the 2D Dirac Hamiltonian:
\begin{equation}\label{Eq16}
    H_\ell^\xi(\hat{\vect{p}}) = - v_F(\xi\sigma_x,\sigma_y)\cdot\left[R^z\left(-\theta_\ell\right)\cdot(\hat{\vect{p}}-\hbar\vect{K}_\xi^\ell)\right],
\end{equation}
where $\sigma_x,\sigma_y$ are two conventional Pauli matrices, $R^z\left(-\theta_\ell\right)$ is a matrix that rotates the relevant vectors around the $Oz$ axis to maintain the canonical form of the Dirac Hamiltonian, $\vect{K}_\xi^\ell$ is the corner point of type (valley) $\xi$ in the first Brillouin zone of layer $\ell$, and $v_F$ is the Fermi velocity. The minus sign in the equation is due to the negative value of the hopping parameter $V_{pp\pi} = -2.7$ eV. The block term for interlayer coupling is expressed as follows:\cite{Koshino_2018}
\begin{align}\label{Eq17}
    T^\xi(\hat{\vect{r}}) =& \left(\begin{array}{cc}
    u & u^\prime \\
    u^\prime    & u 
    \end{array}\right)+\left(\begin{array}{cc}
    u & u^\prime\omega^{-\xi} \\
    u^\prime\omega^{\xi}    & u 
    \end{array}\right)e^{i\xi\delta\vect{k}_2\cdot\hat{\vect{r}}}\nonumber\\
    &+\left(\begin{array}{cc}
    u & u^\prime\omega^{\xi} \\
    u^\prime\omega^{-\xi}    & u 
    \end{array}\right)e^{i\xi\delta\vect{k}_3\cdot\hat{\vect{r}}},
\end{align}
The values of $\delta\vect{k}_2$ and $\delta\vect{k}_3$ are defined as $\vect{A}_1^\star$ and $\vect{A}_1^\star + \vect{A}_2^\star$, respectively, where $\vect{A}_1^\star$ and $\vect{A}_2^\star$ are reciprocal lattice vectors and $\omega = \exp(i2\pi/3)$. To account for the effects of lattice reconstruction, the parameters $u$ and $u^\prime$, which are set to be $0.0797$ eV and $0.0975$ eV, respectively, are chosen based on the description in reference \onlinecite{Koshino_2018}.

The velocity operator corresponding to the Hamiltonian in Eq. (\ref{Eq16}) can be expressed as follows:
\begin{equation}\label{Eq18}
    \hat{v}_\alpha = -v_F(\xi\sigma_x,\sigma_y)\cdot R^z_{\alpha}(-\theta_\ell),
\end{equation}
Here, $R^z_{\alpha}(-\theta_\ell)$ denotes the first or second column of the rotation matrix $R^z(-\theta_\ell)$ depending on whether $\alpha=x$ or $\alpha=y$. It is noteworthy that, because of the relativistic nature of $H^\xi_\ell(\hat{\vect{p}})$ in Eq. (\ref{Eq13}), the current operator $\hat{j}_\alpha$ only comprises a drift current component, $\hat{j}_\alpha = e\hat{v}_\alpha$, with no diffusion or diamagnetic terms.

The spectrum of the Hamiltonian $\hat{H}$ is obtained by solving the secular equation given by:
\begin{align}\label{Eq19}
    \left(\begin{array}{cc}
        H_1^\xi(\hat{\vect{p}}) & T^\xi(\hat{\vect{r}}) \\
    T^{\xi\dagger}(\hat{\vect{r}}) & H_2^\xi(\hat{\vect{p}})
    \end{array}\right)|\psi_{\xi,\vect{k}}\rangle = E|\psi_{\xi,\vect{k}}\rangle,
\end{align}
where the electron states in the twisted bilayer graphene (TBG) lattice are modeled as Bloch state vectors $|\psi_{\xi,\vect{k}}\rangle$. Because of the approximate periodicity of the moire lattice in TBG the state vectors $|\psi_{\xi,\mathbf{k}}\rangle$ can be expanded in terms of the plane-wave vectors $|\vect{k}+\vect{G}_m\rangle$ which are the eigenvectors of the momentum operator $\hat{\vect{p}}$, i.e., $\hat{\vect{p}}|\vect{k}+\vect{G}_m\rangle = (\vect{k}+\vect{G}_m)|\vect{k}+\vect{G}_m\rangle$ and  $\langle\vect{r}|\vect{k}+\vect{G}_m\rangle = e^{i(\vect{k}+\vect{G}_m)\cdot\vect{r}}$. Specifically, we have:
\begin{equation}\label{Eq20}
    |\psi_{\xi,\vect{k}}\rangle = \sum_m\left(\begin{array}{c}
    C_{1,\xi,\vect{k}}(\vect{G}_m) \\
    C_{2,\xi,\vect{k}}(\vect{G}_m)
    \end{array}\right)|\vect{k}+\vect{G}_m\rangle,
\end{equation}
where $C_{1,\xi,\vect{k}}(\vect{G}_m)$ and $C_{2,\xi,\vect{k}}(\vect{G}_m)$ are 2D vectors of combination coefficients that need to be found. $\{\vect{G}_n\,|\, n = 1,2,3, \hdots, N_{\vect{G}}\}$ is a set of $N_{\vect{G}}$ vectors of the moire reciprocal lattice. Substituting the expression (\ref{Eq20}) into Eq. (\ref{Eq19}) then left-multiplying both sides with $\langle\vect{k}+\vect{G}_n|$ we obtain a set of linear equations in the form:
\begin{widetext}
\begin{align}\label{Eq23}
    &\sum_m\left[\left(\begin{array}{cc}
    H_1^\xi(\vect{k}+\vect{G}_n)     & T_1^\xi \\
    T_1^{\xi\dagger}     & H_2^\xi(\vect{k}+\vect{G}_n)
    \end{array}\right)\delta_{\vect{G}_n,\vect{G}_m}+\right.\nonumber\\
    &\hspace{2cm}+\left.\sum_{j=2}^3\left(\begin{array}{cc}
    0  & T_j^\xi \\
    0  & 0
    \end{array}\right)\delta_{\vect{G}_n,\vect{G}_m+\delta\vect{k}_j}+\sum_{j=2}^3\left(\begin{array}{cc}
    0  & 0 \\
    T_j^{\xi\dagger}  & 0
    \end{array}\right)\delta_{\vect{G}_n,\vect{G}_m-\delta\vect{k}_j}\right]\left(\begin{array}{c}
         C_{1,\xi,\vect{k}}(\vect{G}_m)  \\
         C_{2,\xi,\vect{k}}(\vect{G}_m)
    \end{array}\right) = E\left(\begin{array}{c}
         C_{1,\xi,\vect{k}}(\vect{G}_n)  \\
         C_{2,\xi,\vect{k}}(\vect{G}_n)
    \end{array}\right),
\end{align}
\end{widetext}
wherein we denote:
\begin{subequations}
\begin{align}
\langle\vect{k}+\vect{G}_n|H_\ell^\xi|\vect{k}+\vect{G}_m\rangle &= H_\ell^\xi(\vect{k}+\vect{G}_n)\delta_{\vect{G}_n,\vect{G}_m},\label{Eq22a}\\
\langle\vect{k}+\vect{G}_n|T^\xi(\hat{\vect{r}})|\vect{k}+\vect{G}_m\rangle &= \sum_{j=1}^3T_j^\xi\delta_{\vect{G}_n,\vect{G}_m+\delta\vect{k}_j},\label{Eq22b}\\
\langle\vect{k}+\vect{G}_n|T^{\xi\dagger}(\hat{\vect{r}})|\vect{k}+\vect{G}_m\rangle &= \sum_{j=1}^3T_j^{\xi\dagger}\delta_{\vect{G}_n,\vect{G}_m-\delta\vect{k}_j},\label{Eq22c}
\end{align}
\end{subequations}
To obtain a numerical solution, we define a $4N_{\vect{G}} \times 4N_{\vect{G}}$ Hermitian matrix $H_{\vect{k}}^\xi$ in block form for each value of $\xi$ and $\vect{k} \in \text{MBZ}$:
\begin{subequations}
\begin{align}
    &[H_{\vect{k}}^\xi]_{n,n} = \left(\begin{array}{cc}
    H_1^\xi(\vect{k}+\vect{G}_n)     & T_1^\xi \\
    T_1^{\xi\dagger}     & H_2^\xi(\vect{k}+\vect{G}_n)
    \end{array}\right),\label{Eq24a}\\
    &[H_{\vect{k}}^\xi]_{n,m_j} = \left(\begin{array}{cc}
    0     & T_j^\xi \\
    0     & 0
    \end{array}\right), \hspace{0.5cm}\text{if}\hspace{0.5cm} \vect{G}_{m_j} = \vect{G}_n-\delta\vect{k}_j,\label{Eq24b}\\
    &[H_{\vect{k}}^\xi]_{n,m_j^\prime} = \left(\begin{array}{cc}
    0     & 0 \\
    T_j^{\xi\dagger}     & 0
    \end{array}\right), \hspace{0.4cm}\text{if}\hspace{0.5cm} \vect{G}_{m_j^\prime} = \vect{G}_n+\delta\vect{k}_j.\label{Eq24c}
\end{align}
\end{subequations}
The eigenvalues $E_n^{\xi}(\vect{k})$ and corresponding eigenvectors, obtained by diagonalizing the matrix $H_{\vect{k}}^\xi$, approximate the states of low-energy electrons in the TBG lattice. The accuracy of the effective description is numerically dependent on the number of reciprocal lattice vectors $\{\vect{G}_n\,|\, n = 1,2,3,\hdots, N_{\vect{G}}\}$ considered in the calculation. To ensure validity in the low-energy range near the intrinsic Fermi level, where the energy surfaces of the monolayer graphene take the form of cones, we control the value of $N_{\vect{G}}$ by using a cutoff energy $E_c$. Specifically, we count only the $\vect{G}_n$ vectors such that $|\vect{G}_n|\leq E_c/\hbar v_F$.
\begin{figure}\centering
\includegraphics[clip=true,trim=2cm 11.5cm 5cm 7cm,width=\columnwidth]{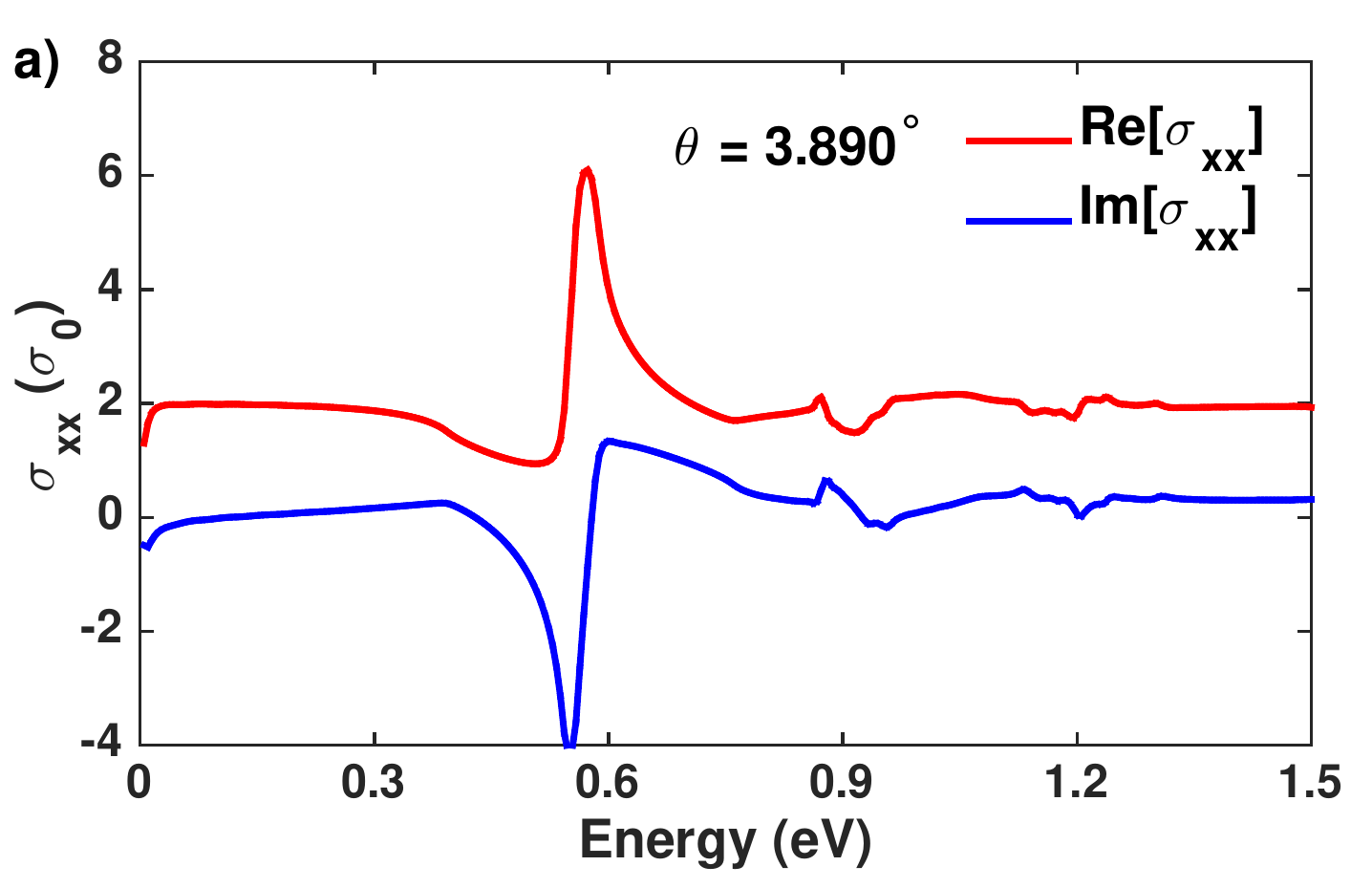}\\
\includegraphics[clip=true,trim=2cm 11.5cm 5cm 7cm,width=\columnwidth]{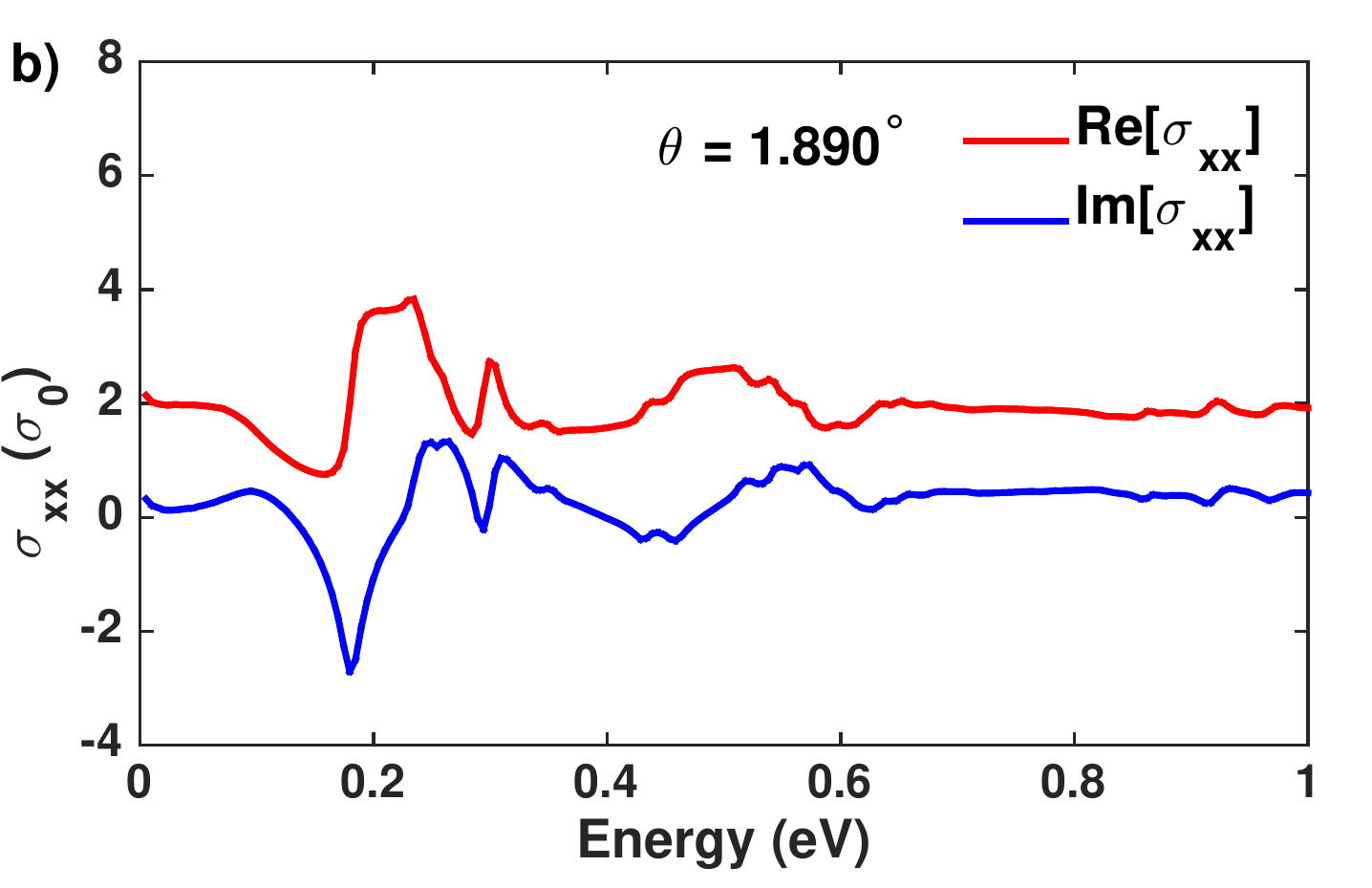}\\
\includegraphics[clip=true,trim=2cm 11.5cm 5cm 7cm,width=\columnwidth]{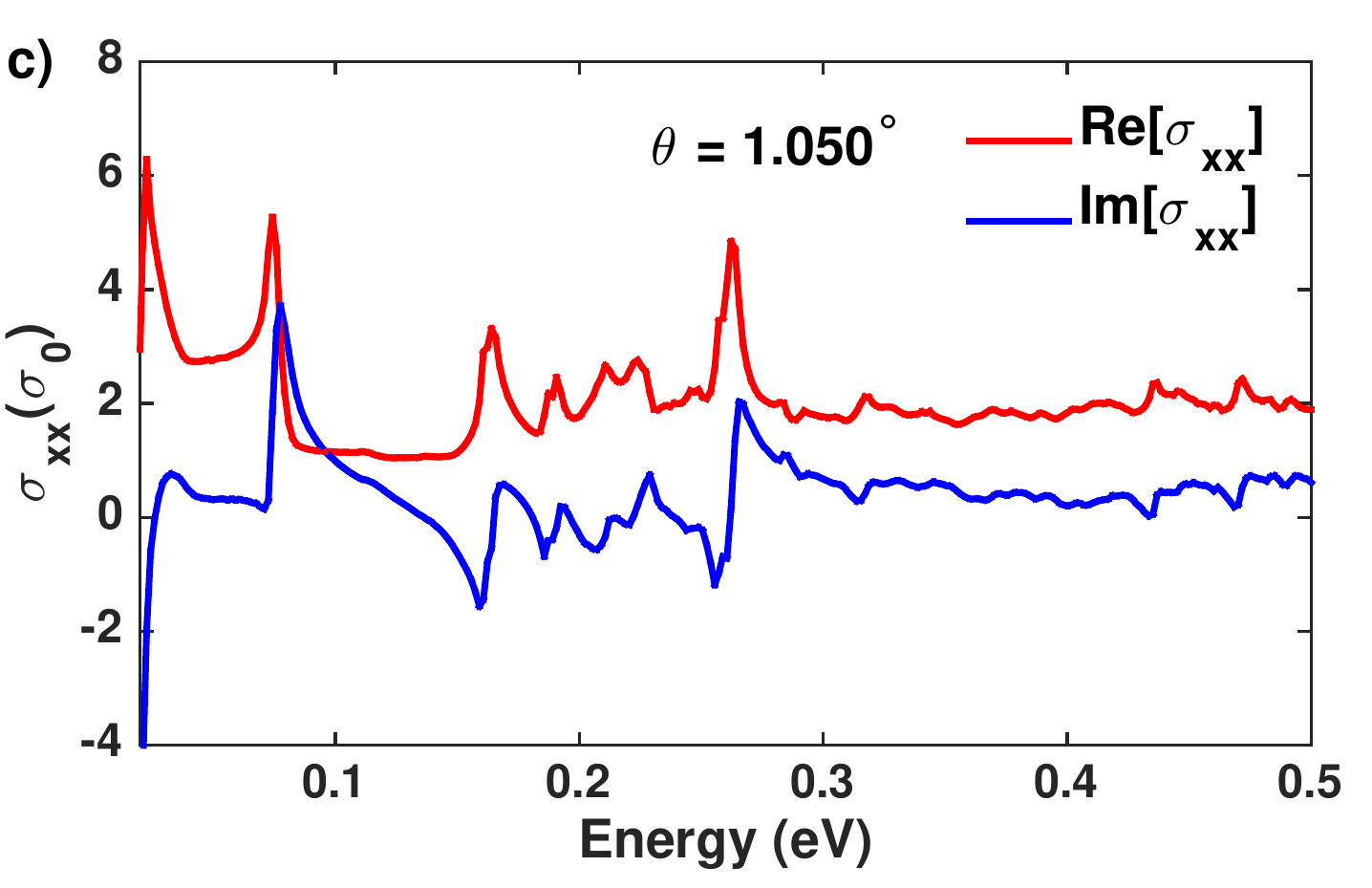}
\caption{\label{Fig3} The real and imaginary components of the overall longitudinal conductivity for TBG configurations with $\theta = 3.890^\circ$ (a), $1.890^\circ$ (b), and $1.050^\circ$ (c). There are no total transverse conductivities due to the presence of time-reversal symmetry.}
\end{figure}
\begin{figure}\centering
\includegraphics[clip=true,trim=1.6cm 11.5cm 5cm 7cm,width=\columnwidth]{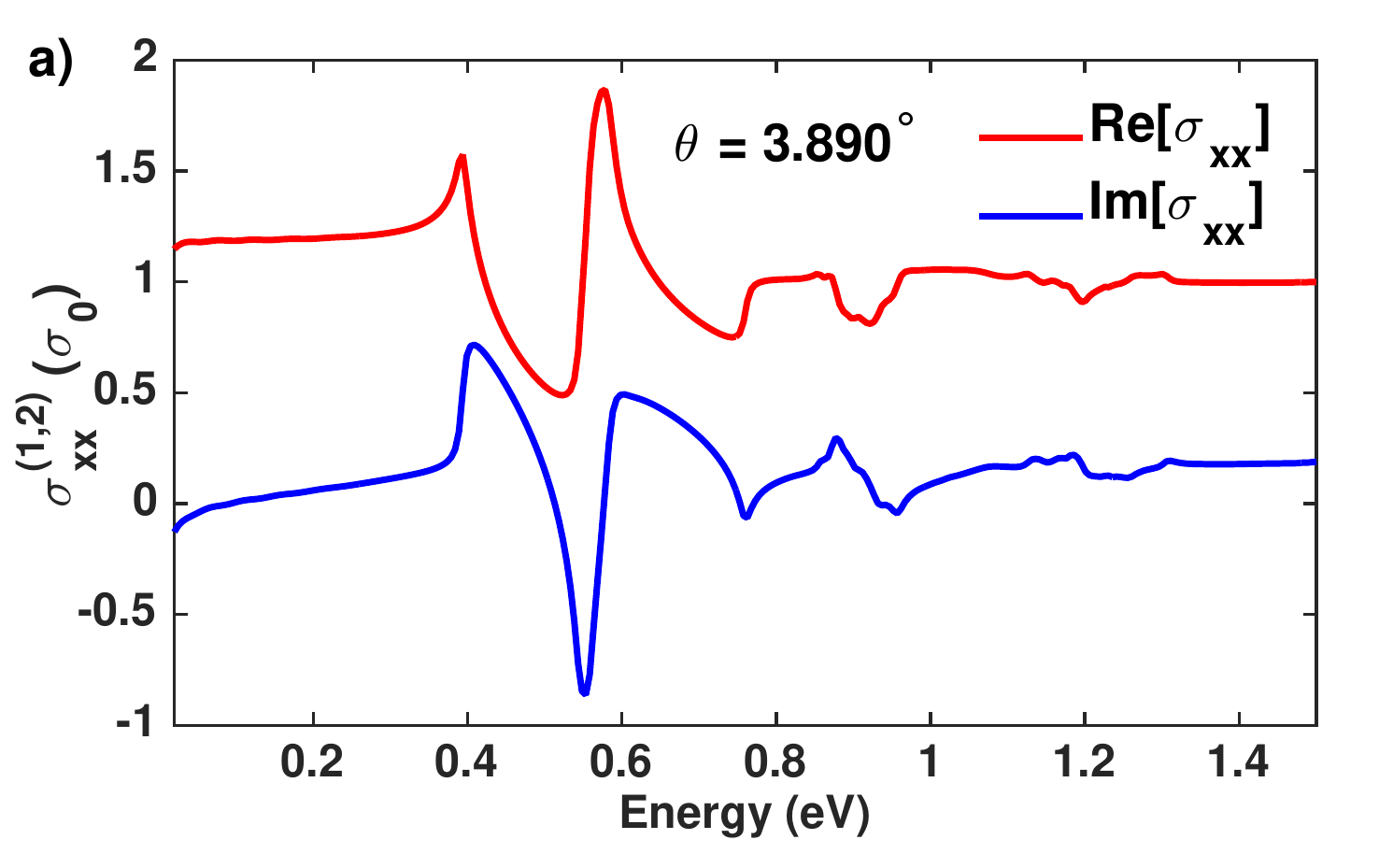}\\
\includegraphics[clip=true,trim=1.6cm 11.5cm 5cm 7cm,width=\columnwidth]{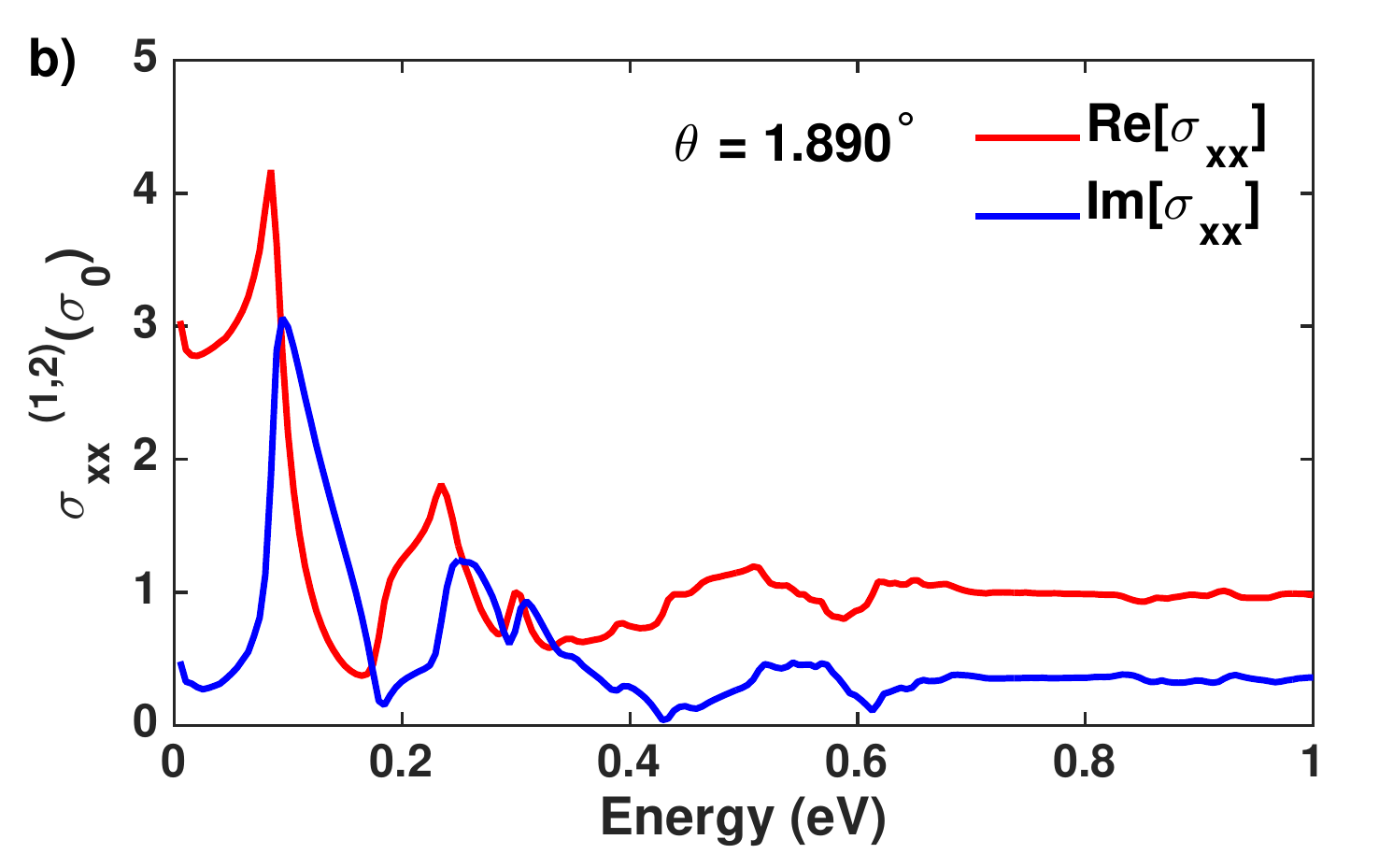}
\\
\includegraphics[clip=true,trim=1.6cm 11.5cm 5cm 7cm,width=\columnwidth]{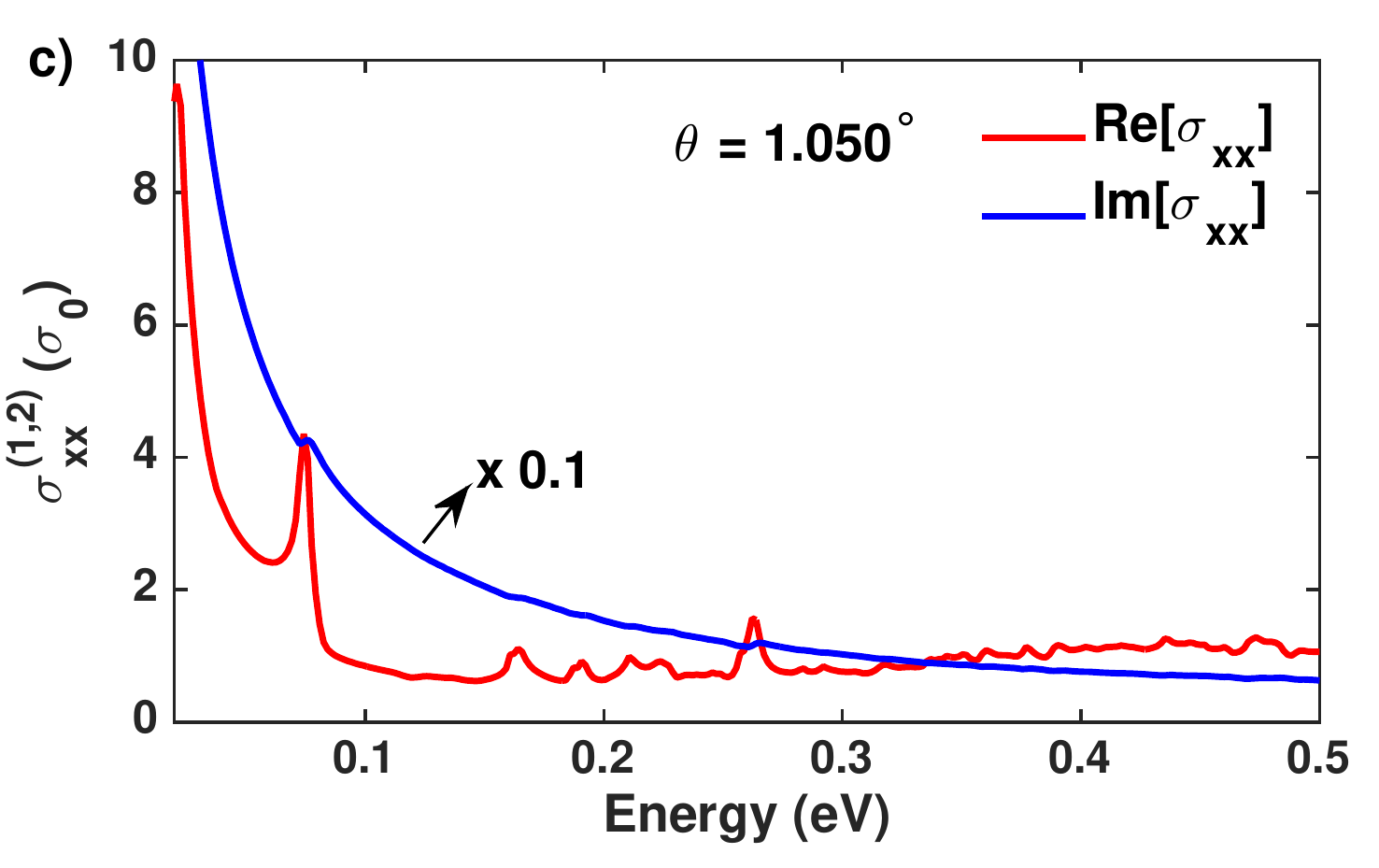}
\caption{\label{Fig4} The real and imaginary parts of the diagonal element $\sigma_{xx}^{(\ell)}$ of the tensor $\boldsymbol{\sigma}^{(\ell)}(\omega)$ ($\ell = 1, 2$) for three TBG configurations with $\theta = 3.890^\circ, 1.890^\circ$ and $1.050^\circ$.}
\end{figure}
\section{Results and discussions}\label{SecIII}
\subsection{Electronic band structure of TBGs}\label{SecIIIA}
We used numerical methods to solve Eq. (\ref{Eq23}). Specifically, for each value of $\xi=\pm 1$, we constructed a Hermitian matrix $H_{\vect{k}}^\xi$ of size $4N_{\vect{G}}\times 4N_{\vect{G}}$ and diagonalized it for each value of $\vect{k}$ in the mini-Brillouin zone of the TBG system. The sets of eigenvalues ${E_m(\vect{k})}$ and eigenvectors $\{\mathbf{C}^m_{\xi}(\vect{k})=[C^m_{1,\xi,\vect{k}}(\vect{G}_1),C^m_{2,\xi,\vect{k}}(\vect{G}_1),\hdots,C^m_{1,\xi,\vect{k}}(\vect{G}_{N_{\vect{G}}}),C^m_{2,\xi,\vect{k}}(\vect{G}_{N_{\vect{G}}})]^T\}$ were calculated. The results for three TBG configurations with twist angles of $\theta = 1.05^\circ, 1.89^\circ$ and $3.89^\circ$ are shown in Fig. \ref{Fig1}. The blue curves represent the results obtained when the electronic interlayer coupling is artificially turned off. Both sets of data are plotted on the same figure for comparison, to highlight the impact of the interlayer coupling. The presence of multiple dispersion curves in the figure results from the folding of the energy band structure of two graphene layers due to the enlargement of the unit cell of the TBG lattice and the corresponding shrinkage of the Brillouin zone into the mini-Brillouin zone. However, the electronic band structure of the TBG system is not simply formed in this manner. At the points where the bands of monolayer graphene cross, the interlayer coupling leads to the hybridization of Bloch states and the lifting of energy degeneracy to form new bands. Our results agree with other available data in the literature \cite{Morell_2010,Laissardiere_2012,Moon-2013,Koshino_2018}. However, in this study, we wish to emphasize the formation of new electronic states, which are special to TBG systems and play a crucial role in determining its electronic, optical and transport properties. Fig. \ref{Fig1} can be considered a visual representation of the evolution of the band structure with respect to the twist angle. We observe that for the configuration with $\theta = 3.89^\circ$, the dispersion curves in the low energy range around the intrinsic Fermi energy ($E_F = 0$) have a similar form to the blue curves, although they are not identical. This implies that the states created in the bilayer system share similarities with those in monolayer graphene. For smaller twist angles, particularly $\theta = 1.05^\circ$, a band with a very narrow bandwidth forms around the Fermi energy, separated from the lower and upper bands by a narrow gap. These dispersion curves are completely different from the blue curves, as seen in Fig. \ref{Fig1}(e). We conclude that the electronic interlayer coupling creates special states in TBG lattices with small twist angles. Although the TBG systems are two-dimensional, it can be challenging to analyze their geometrical features of energy surfaces in wave-vector space. However, the density of states (DOS) generally exhibits key features, such as van Hove singularity behavior, as shown in the right panels of Figs. \ref{Fig1} (a, b, c). The blue curves in these figures were obtained numerically without considering the effects of electronic interlayer coupling, which is equivalent to the DOS of two independent graphene layers. The red curves show significant peaks due to van Hove singularities, indicating the presence of extremal and saddle points of the energy surfaces.

It's worth noting that the energy bands with index $\xi =+1$ (resulting from the hybridization of graphene Bloch states in the $K_1$ valley of the first layer and $K_2$ valley of the second layer) differ from those with index $\xi = -1$ (resulting from the hybridization of Bloch states in the $K_1^\prime$ valley and $K_2^\prime$). This difference is evident along the ${\it\Gamma}M$ direction as shown in Figs. \ref{Fig1} (d, e, f). However, it does not appear in the DOS picture. We computed the DOS for both $\xi = +1$ and $\xi = -1$ separately and obtained the same results. This implies that the energy surfaces with $\xi = +1$ are simply rotations of the corresponding ones with $\xi = -1$ by an angle of $\pi/3$.
\begin{figure*}\centering
\includegraphics[clip=true,trim=1.6cm 11cm 5cm 7cm,width=0.45\textwidth]{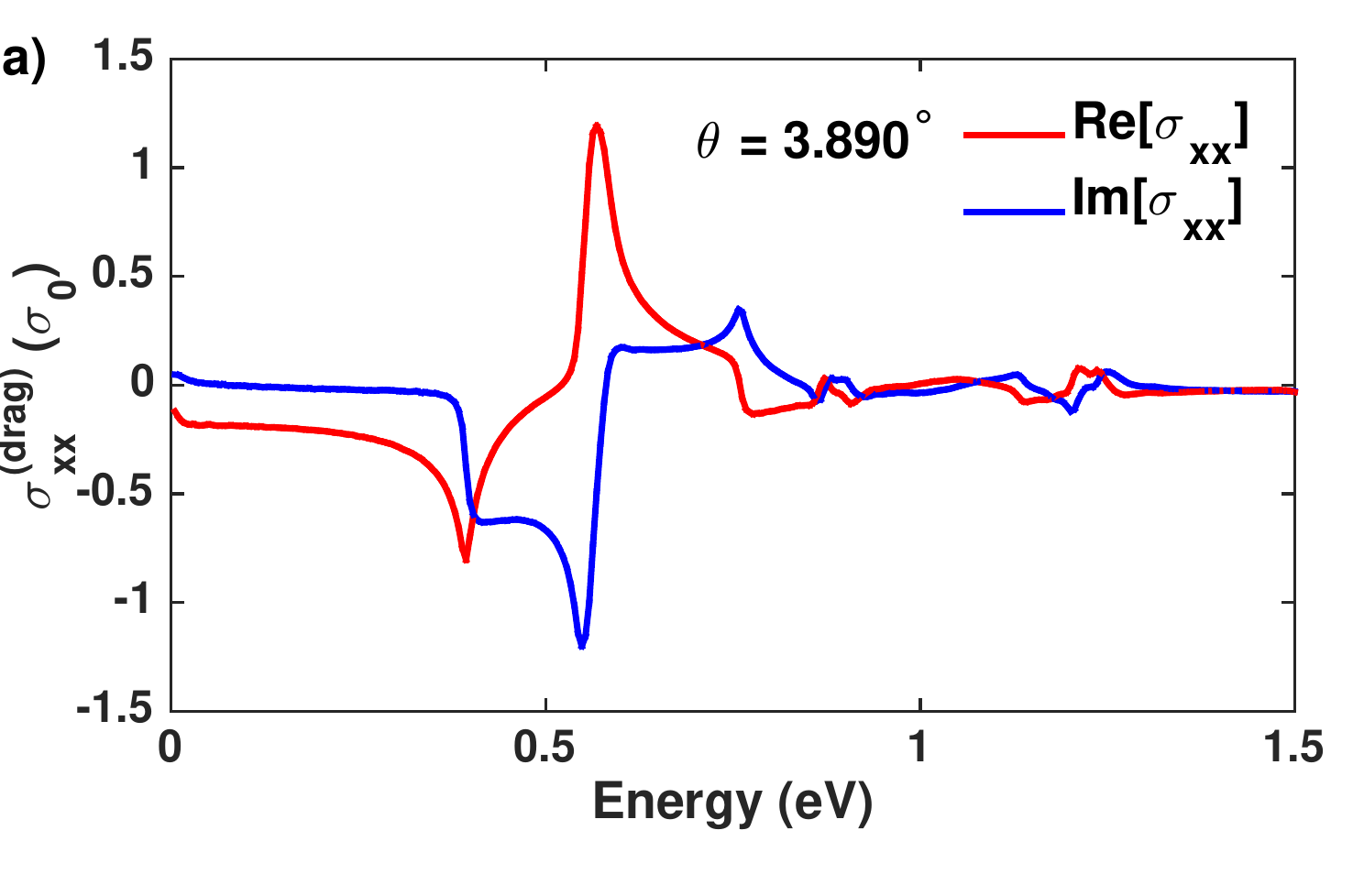}
\includegraphics[clip=true,trim=1.3cm 11cm 5.3cm 7cm,width=0.45\textwidth]{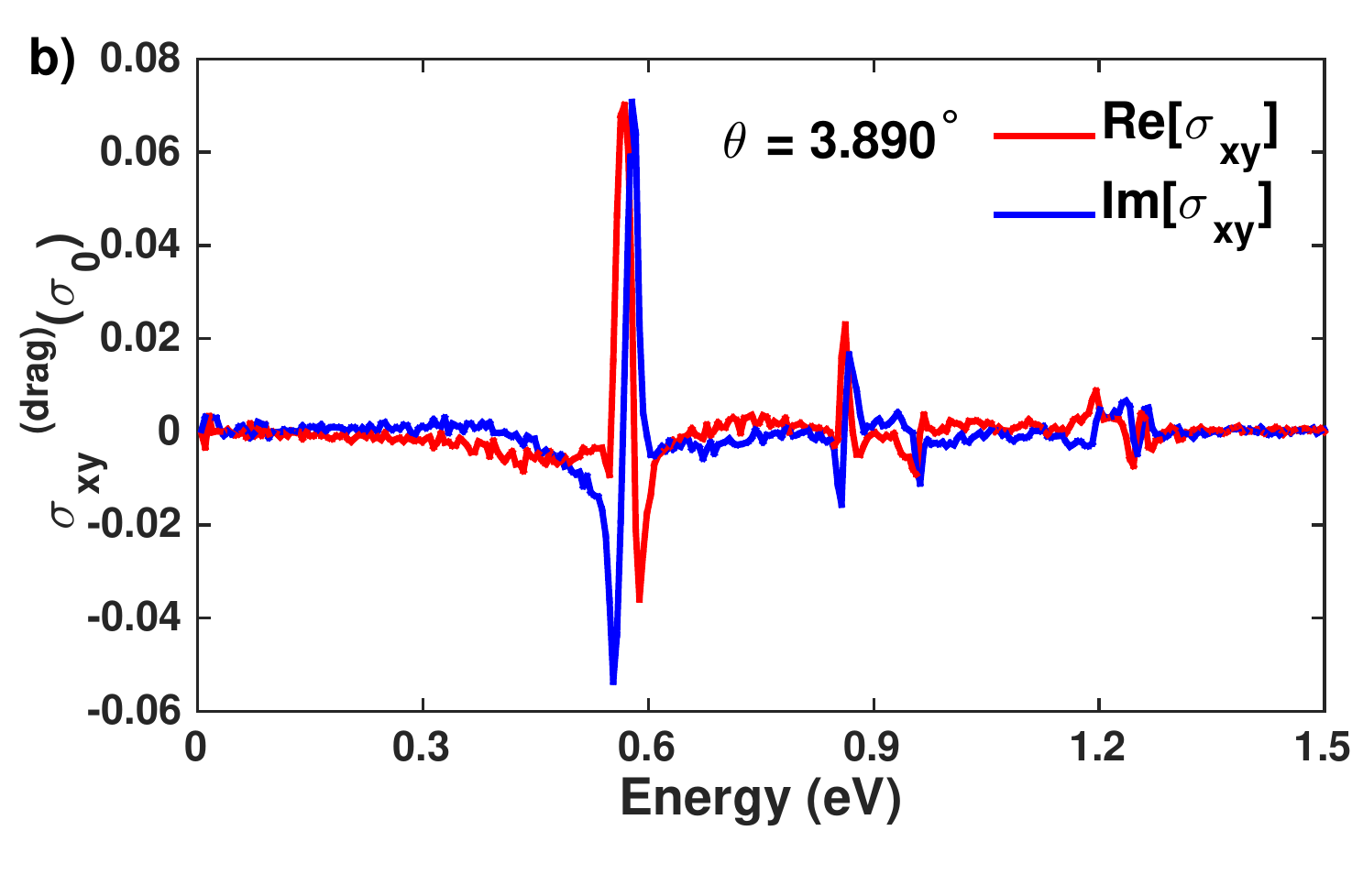}\\
\includegraphics[clip=true,trim=1.6cm 11cm 5cm 7cm,width=0.45\textwidth]{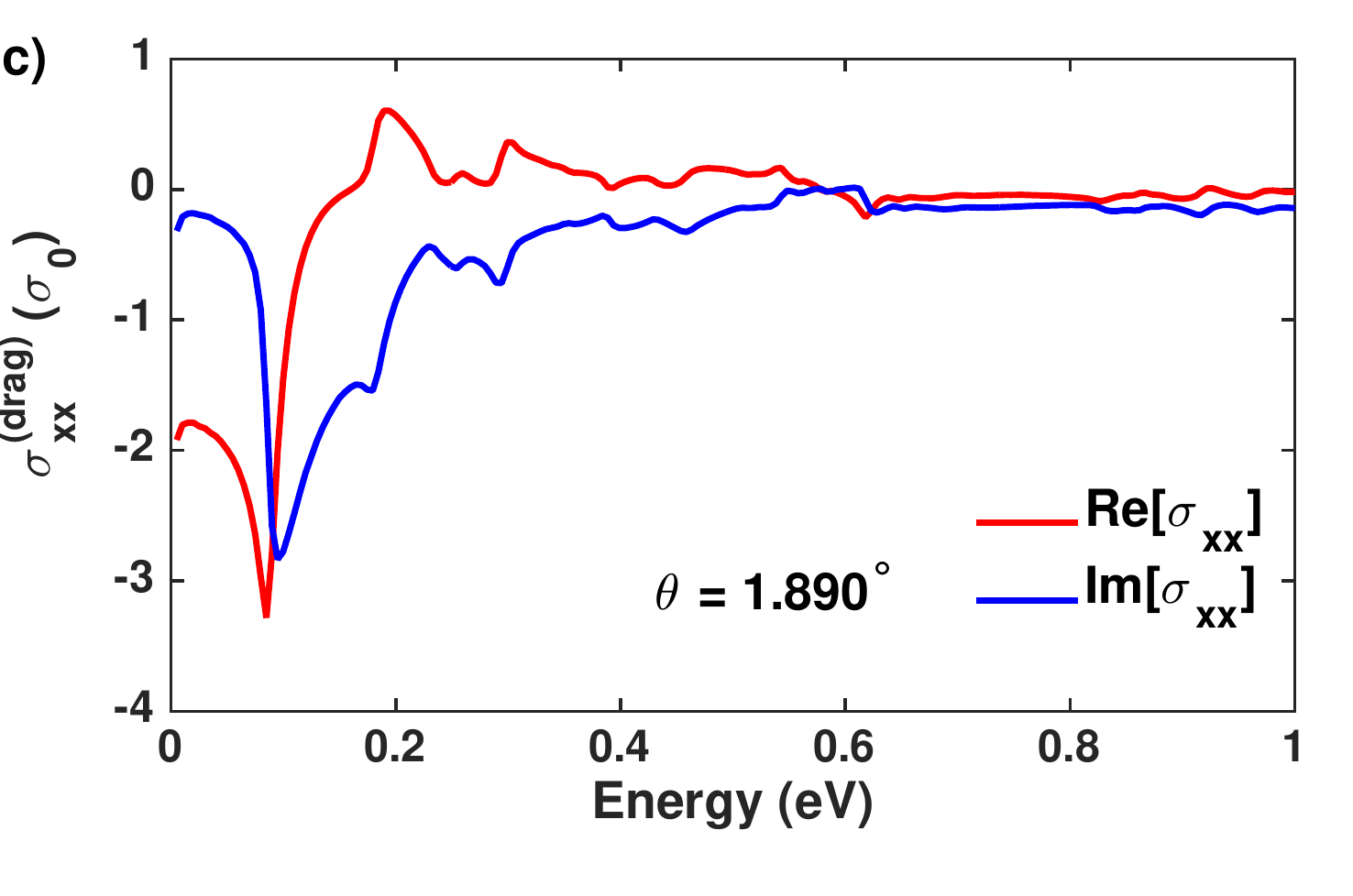}
\includegraphics[clip=true,trim=1.3cm 11cm 5.3cm 7cm,width=0.45\textwidth]{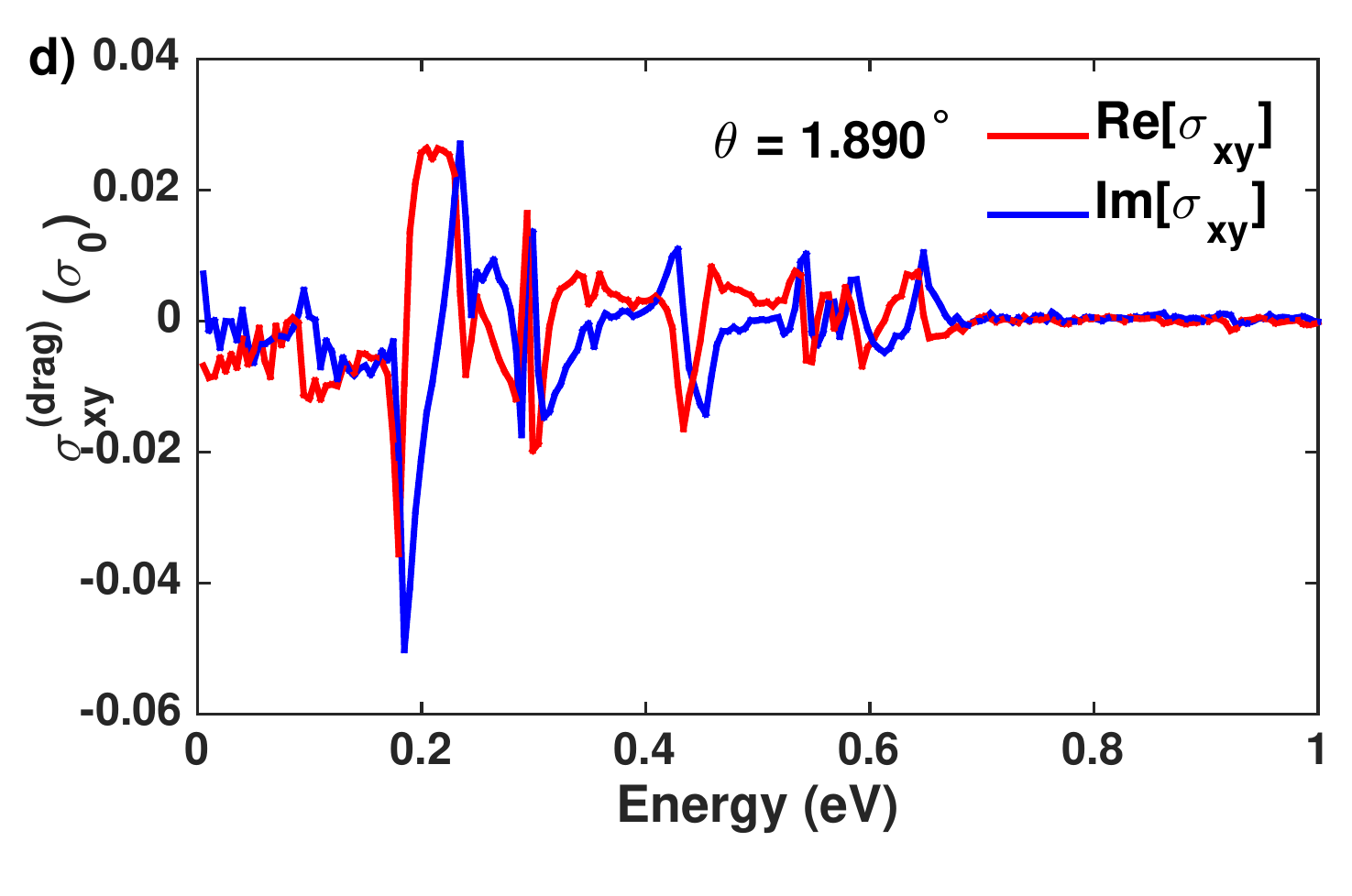}\\
\includegraphics[clip=true,trim=1.6cm 11cm 5cm 7cm,width=0.45\textwidth]{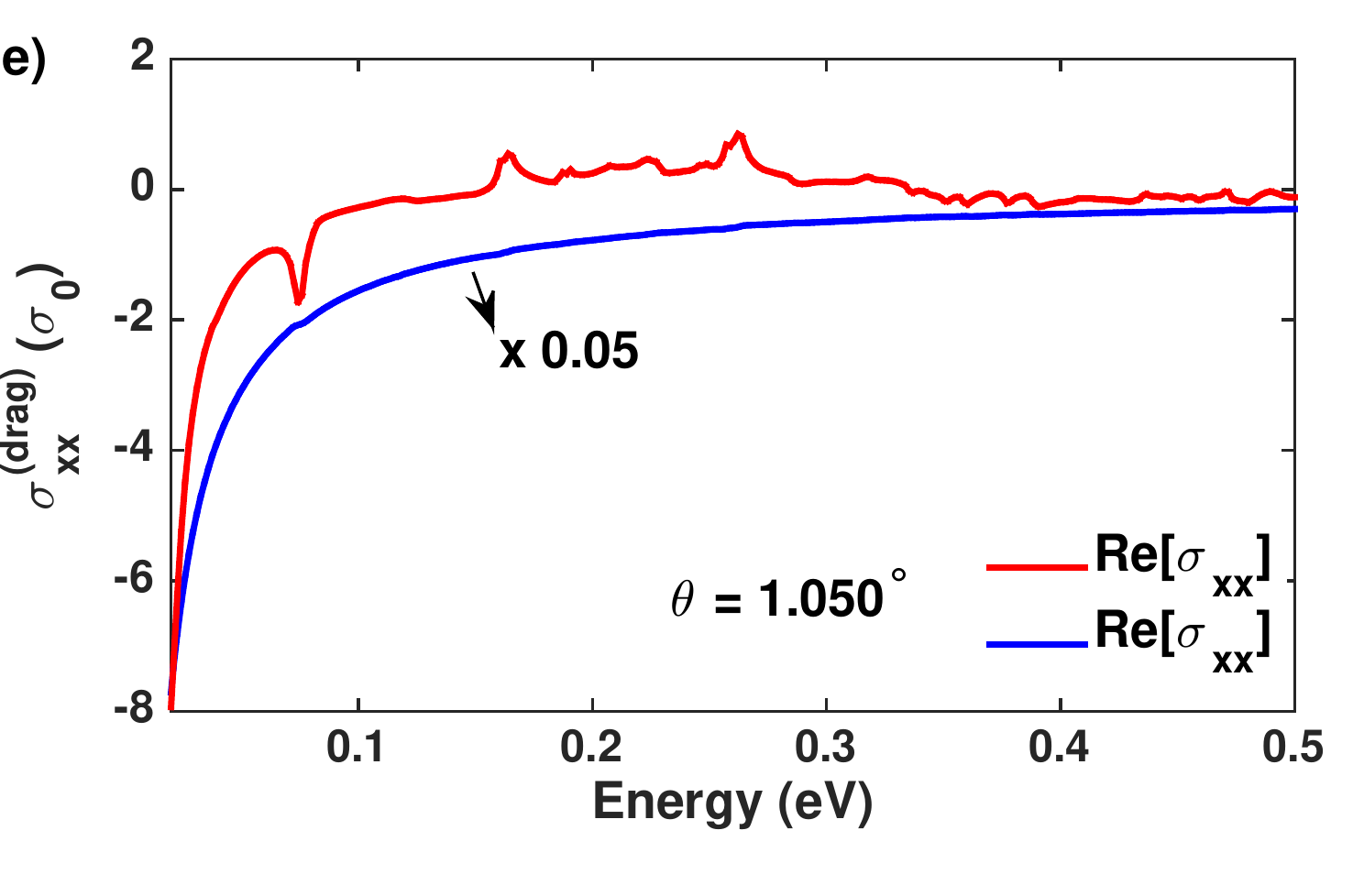}
\includegraphics[clip=true,trim=1.3cm 11cm 5.3cm 7cm,width=0.45\textwidth]{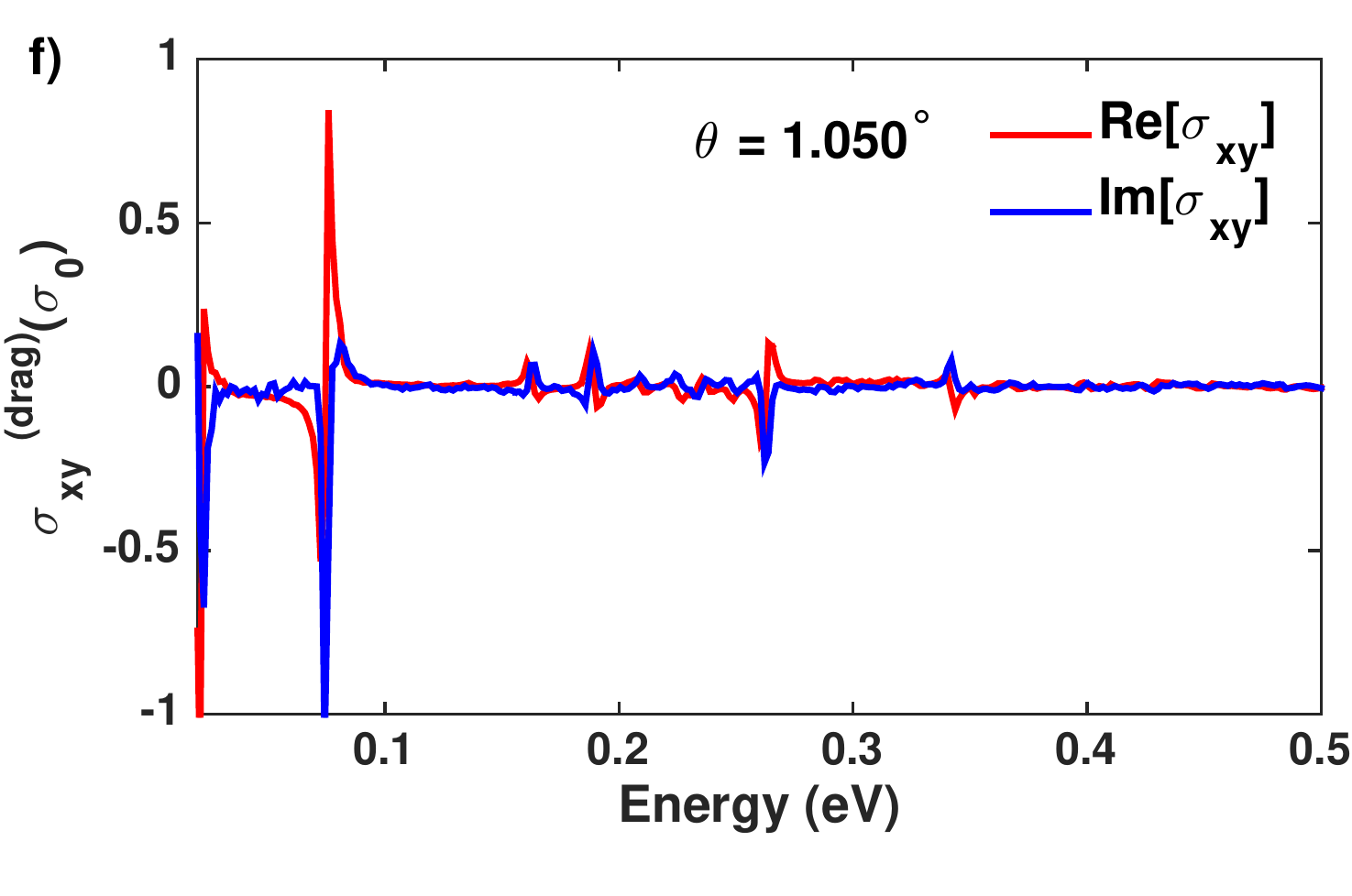}
\caption{\label{Fig5} The real and imaginary parts of the drag component of the conductivity tensor, $\boldsymbol{\sigma}^{(drag)}(\omega)$, for three TBG configurations with $\theta = 3.890^\circ, 1.890^\circ$, and $1.050^\circ$. (a,c,e) represent the longitudinal (diagonal) component $\sigma_{xx}^{(drag)}(\omega)$, and (b,d,f) represent the transverse (off-diagonal) component $\sigma_{xy}^{(drag)}(\omega)$.}
\end{figure*}
\subsection{DC conductivity}\label{SecIIIB}
The calculation of the electronic structures of TBG configurations with twist angles greater than 2 degrees shows the presence of two energy bands near the Fermi energy ($E_F = 0$), which have dispersion curves similar to those of monolayer graphene. However, the Fermi velocity of the linear dispersion law in TBG ($v_F^{TBG}$) is smaller than that of monolayer graphene ($v_F^{MLG}$). As the twist angle decreases, these two bands become closer and eventually form an isolated band with a narrow bandwidth. This ``flat band" near the Fermi energy is expected to have limited impact on the transport properties of TBG sheets because of its small Fermi velocity. In other words, electrons occupying these bands are expected to be spatially confined in the atomic lattice. Although these bands are not completely dispersionless, the study of their transport properties is still of great interest. To this end, within the framework of the non-interacting electron approximation, the DC conductivity was calculated using the Kubo-Greenwood formula as given by Eqs. (\ref{Eq9}) and (\ref{Eq10}). The results of the calculation for the longitudinal components ($\sigma_{xx}^{DC}$ and $\sigma_{yy}^{DC}$) of three TBG configurations at low temperatures are presented in Fig. \ref{Fig2}.

The data indicate that the electrical conductivity of TBG systems is anisotropic, which is reflected in the two different longitudinal elements of the conductivity tensor, $\sigma_{xx}^{DC}$ (represented by solid curves) and $\sigma_{yy}^{DC}$ (represented by dashed curves). The conductivity should reflect the electronic structure with respect to the Fermi energy, and our calculations for the two cases of $\xi = \pm 1$ yielded the same result. This is similar to the picture of the density of states, but the difference between the two longitudinal conductivities highlights the anisotropic nature of the energy surfaces. For the TBG configuration with a twist angle of $\theta = 3.890^\circ$, the electronic structure is similar to that of monolayer graphene (except for the value of the Fermi velocity), and the conductivity curves have a characteristic V-shape (or an M-shape when considered over a wider energy range, as shown by the blue curves). For TBG configurations with smaller twist angles, the electronic structure and thus the conductivity picture becomes much more complex. However, our calculations show that the conductivity at the intrinsic Fermi level $E_F = 0$ is always finite and non-zero. Interestingly, the calculated DC conductivity for TBG configurations with twist angles that are not magic values is about two times the quantum value of the minimal conductivity of monolayer graphene, which is $\sigma^{DC}_0 = 4e^2/\pi h$. In the case of the TBG configuration with the magic twist angle $\theta = 1.050^\circ$, the conductivity curve shows a small hump at the Fermi energy $E_F = 0$ with $\sigma_{\alpha\alpha}^{DC}(0) \approx \sigma^{DC}_0$. This quantum value of the conductivity of monolayer graphene has been intensively discussed in the past and is rooted in the fact that the valence and conduction bands touch each other at the $K$ points, resulting in always-available free carriers due to fluctuations in the electrostatic profile around the Fermi energy level.\cite{Novoselov_2005, Katsnelson_2006, Tworzydlo_2006, Miao_2007, Blake_2009, Do_2010} In our calculation, we approximated the delta-Dirac functions in the Kubo-Greenwood formula with a Gaussian function with a parameter $\eta$ that defines the finite width of the peak. The value of $\eta$ was taken to be $\eta < 5$ meV to account for both numerical calculation needs and the broadening of energy levels and finite lifetime of the electric charge-carrying quasi-particles. Although the observed DC conductivity value in TBG systems is supported by the characteristic hybridized states in the TBG lattice, 
we suppose that it has the same underlying physical explanation as in monolayer graphene.
\subsection{AC conductivity and optical Hall drag conductivity}\label{SecIIIC}
In Fig. \ref{Fig3}, we present the results of our calculations for the total optical conductivity of three different twisted bilayer graphene (TBG) configurations. The calculation was performed using the Kubo formula for the conductivity tensor, where the velocity operator $\hat{v}_\alpha$ was determined from the Hamiltonian (\ref{Eq15}) using Eq. (\ref{Eq6}). The time reversal symmetry and the Onsager reciprocal relations eliminate the off-diagonal elements $\sigma_{xy}$ and $\sigma_{yx}$ of the conductivity tensor, which we numerically confirmed behave as noise with an amplitude of $10^{-3}$. We also verified all of the symmetry properties of the conductivity tensor according to Eq. (\ref{Eq5}), demonstrating the accuracy of our numerical data.

Figure \ref{Fig3} displays the longitudinal conductivity curves $\sigma_{xx}(\omega)$ with significant peaks, which are attributed to dominant interband transition processes. This correlation is supported by its relationship to the electronic band structures of the TBG configurations, shown in Fig. \ref{Fig1}. For instance, in Fig. \ref{Fig1}(b), at an energy of 0.5681 eV, there is an optical absorption peak that can be assigned to a transition from the highest valence band to the lowest conduction band around the $M$ points in the mini-Brillouin zone. As the twist angle decreases, the electronic structure becomes more complex and the optical conductivity exhibits more optical absorption peaks.

For TBG configurations with twist angles greater than 2 degrees, there is a distinctive feature in the low-energy range of the conductivity curve: the real part of the conductivity is independent of the photon energy $\hbar\omega$ and takes a value of $2\sigma_0$, where $\sigma_0 = \pi e^2/2h$. This behavior is similar to that of monolayer graphene and results from the linear dispersion law of electron states. In the low-energy range, the energy band structure of TBG systems is qualitatively similar to that of monolayer graphene, but with a smaller Fermi velocity, as previously analyzed. The value of $\sigma_{xx}(\omega)$ is equal to $2\sigma_0$, as found in previous studies.\cite{Do_2018,Do_2020,Moon-2013}

For the case of the ``magic angle" $\theta = 1.050^\circ$, our calculation shows a clear dominant contribution from optical transitions from the quasi-flat bands around the Fermi level to the lowest conduction band around the $\Gamma$ point, as shown in Fig. \ref{Fig3}(c) and Figs. \ref{Fig1}(c) and \ref{Fig1}(f).

Thus far, we would like to emphasize that the use of the total DC conductivity tensor is not relevant when studying the wave transmission problem in TBG. As we have presented in Secs. \ref{SecIIA} and \ref{SecIIB}, the required quantities to be utilized are $\boldsymbol{\sigma}^{(\ell)}(\omega)$ and $\boldsymbol{\sigma}^{(drag)}(\omega)$. In Figs. \ref{Fig4} and \ref{Fig5}, we present the calculation results for the elements of the conductivity tensor blocks $\boldsymbol{\sigma}^{(\ell)}(\omega)$ and $\boldsymbol{\sigma}^{(drag)}(\omega)$ as analyzed in Sec. \ref{SecIIB}. Our numerical calculation reveals that $\boldsymbol{\sigma}^{(1)}(\omega) = \boldsymbol{\sigma}^{(2)}(\omega) = \sigma_{xx}^{(\ell)}(\omega)\vect{\tau}_0$, verifying the equivalent role of the two graphene layers in the TBG lattice. Figs. \ref{Fig4}(a), (b), and (c) display the real (red) and imaginary (blue) parts of $\sigma_{xx}^{(\ell)}(\omega)$ for $\ell=1,2$. It is worth noting that decreasing the twist angle increases the magnitude of $\sigma_{xx}^{(\ell)}$, particularly when the imaginary part becomes dominant and varies in accordance with $1/\hbar\omega$, as seen in Fig. \ref{Fig4}(c). Fig. \ref{Fig5} shows the real and imaginary parts of the off-diagonal element $\boldsymbol{\sigma}^{(drag)}(\omega)$ of the drag part of the conductivity tensor. For large twist angles in the TBG configuration, there is a notable difference between $\sigma_{xx}^{(\ell)}(\omega)$ and $\sigma_{xx}^{(drag)}(\omega)$, see Figs. \ref{Fig4}(a) and \ref{Fig5}(a). However, for small twist angles ($<2^\circ$), these curves are quite similar but have opposite signs. As a result, although there is significant variation in the magnitude of $\sigma_{xx}^{(\ell)}(\omega)$ and $\sigma_{xx}^{(drag)}(\omega)$, they compensate each other when summed to give the total value $\sigma_{xx}(\omega) = 2[\sigma_{xx}^{(\ell)}(\omega)+\sigma_{xx}^{(drag)}(\omega)]$, as seen in Fig. \ref{Fig2}. 


Figures \ref{Fig5}(b), (d), and (f) present the calculated results for the transverse conductivity component $\sigma_{xy}^{(drag)}(\omega)$. The behavior of $\sigma_{xy}^{(drag)}(\omega)$ is primarily characterized by close-to-zero values across the energy range, with some sharp peaks and dips appearing at energy positions with high optical absorption. This finite value of $\sigma_{xy}^{(drag)}(\omega)$ can be attributed to both the correlation between the current densities in the two graphene layers and the chiral structure of the atomic lattices. The former is considered a necessary condition, representing the result of the coupling between the two graphene layers, which leads to the formation of composite states like the superposition of Bloch states between the two layers. This means that when electrons move in one graphene layer along the $Ox$ direction, they induce a motion in the second layer along the transverse $Oy$ direction due to interlayer coupling. The sufficient condition, however, is the contribution of the electronic states in the system to the resulting conductivity. If the system had mirror symmetry, these correlations would cancel each other out completely. However, the TBG lattices lack mirror symmetry, unlike the hexagonal lattice of graphene, which results in a finite, non-zero conductivity of $\sigma_{xy}^{(drag)}(\omega) \neq 0$. 

\begin{figure}\centering
\includegraphics[clip=true,trim=1.5cm 6.5cm 2.5cm 7cm,width=\columnwidth]{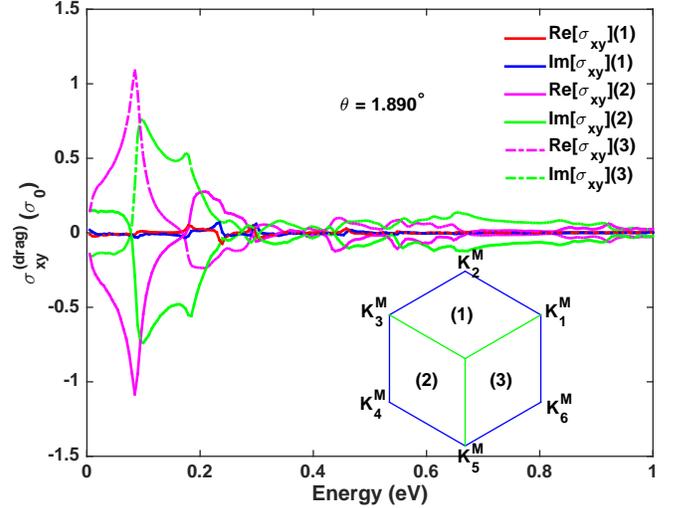}
\caption{\label{Fig6} Contribution of Bloch States to $\sigma_{xy}^{(drag)}(\omega)$: The red and blue curves represent the contribution of Bloch states with $\vect{k}$ in the first third of the Brillouin zone, while the solid purple and dashed green curves represent the contribution of Bloch states with $\vect{k}$ in the second and third thirds of the Brillouin zone, respectively.}
\end{figure}

To emphasize this point, the hexagonal mini-Brillouin zone of the TBG lattice was divided into three equal parts, as depicted in the inset of Fig. \ref{Fig6}. This division was based on the assumption of mirror symmetry through the plane $M_{yz}$, with domain (1) containing the symmetry, and domains (2) and (3) being its mirror images. Using the Kubo formula, the value of $\sigma_{xy}^{(drag)}(\omega)$ was calculated by summing over all $\vect{k}$ points in each of the three domains. The results, shown in Fig. \ref{Fig6}, indicate that the results in domains (2) and (3) are similar, opposite in sign, and much larger than the result in domain (1). Upon summing these results, a strong but incomplete cancellation takes place, leading to the final value of $\sigma_{xy}^{(drag)}(\omega)$ displayed in Figs. \ref{Fig5}(b), (d), and (f). For the AA- and AB-stacked configurations with mirror symmetries, the cancellation was found to be complete, resulting in $\sigma_{xy}^{(drag)}(\omega) = 0$. The symmetry analyses in Refs. \onlinecite{Morell_2010} and \onlinecite{Do_2020} support this theory.

\begin{figure}\centering
\includegraphics[clip=true,trim=1.5cm 6.5cm 2.5cm 7cm,width=\columnwidth]{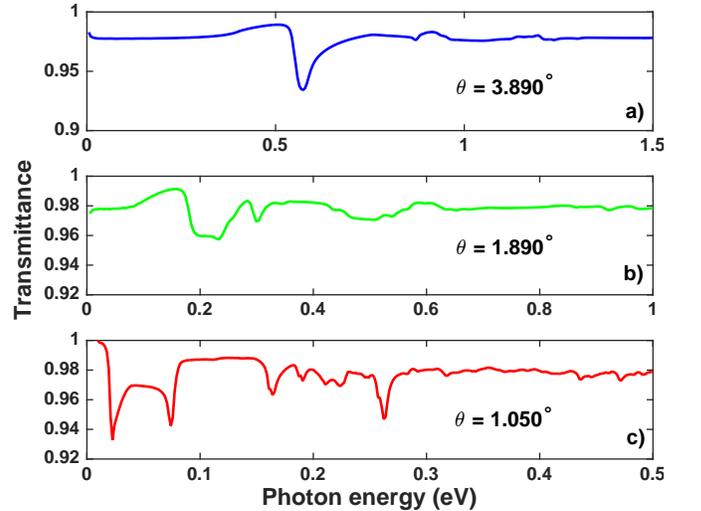}
\caption{\label{Fig7} The transmission spectra for three TBG configurations are shown in (a), (b), and (c) respectively, with twist angles of $\theta = 3.890^\circ$, $\theta = 1.890^\circ$, and $\theta = 1.050^\circ$.}
\end{figure}

\begin{figure}\centering
\includegraphics[clip=true,trim=1.3cm 6.5cm 2.5cm 7cm,width=\columnwidth]{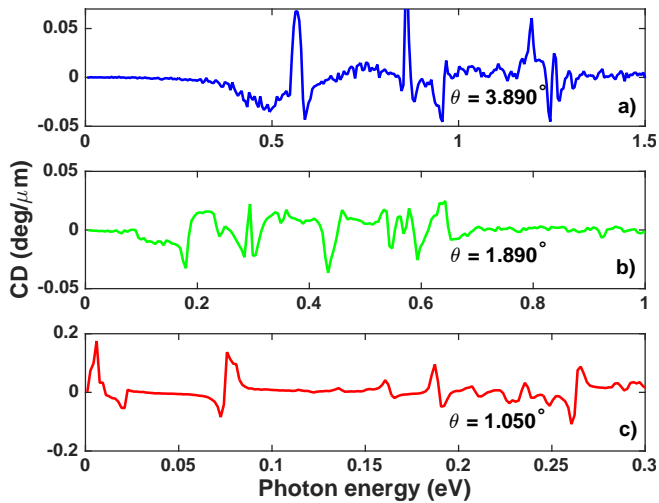}
\caption{\label{Fig8} The circular dichroism (CD) spectra of three TBG configurations: with twist angles of $\theta = 3.890^\circ$ (a), $\theta = 1.890^\circ$ (b), and $\theta = 1.050^\circ$ (c) are displayed.}
\end{figure}

\subsection{Optical characteristics}\label{SecIIID}
On the basis of the theory established in Sec. \ref{SecII} and the data of the conductivity tensor already computed, we performed the calculation for the transfer matrix, then the reflection, transmission and absorption spectra of several TBG configurations.  The obtained results for transmittance and CD are presented in Figs. \ref{Fig7} and \ref{Fig8}, respectively. Since the reflectance is negligible, it is not shown here. It should be noted that while the optical properties of TBGs have been extensively studied, they are usually analyzed through the behavior of the real and imaginary parts of the diagonal components of the total optical conductivity tensor,\cite{Nicol_2008,Moon-2013,Catarina_2019} rather than through the transmittance and reflectance, which are quantities that can be directly measured in experiments.  From the transmission spectra as shown, we see that the transmittance of all TBG configurations can get the value of 98\% and the absorptance of 2\% in average in a large energy range. The spectral curves clearly exhibit the peaks and dips consistent with the picture of the longitudinal conductivities. The transmission and absorption spectra for the left- and right-handed lights are different to be distinguished. However, the CD spectra is clearly manifested. From Fig. \ref{Fig8}, we see that for the TBG configuration with $\theta = 3.890^\circ$,  the CD curve exhibits essential features of experimental data reported by Kim et. al.,\cite{Kim_2016} for instance, with the peaks and dips of large width, except several sharp peaks at the energy position of the absorption peaks. In order to verify the role of the drag transverse conductivity $\sigma_{xy}^{(drag)}$, we did not include it into the calculation. As expected, it results in the zero CD for the total energy range, but mostly does not change the transmission spectrum. We therefore conclude that this drag transverse conductivity plays the decisive role in governing the chiral optical behaviors of the TBG systems. Last but not least, it is important to note that the CD spectrum shown in Fig. \ref{Fig8} is extracted from the theory presented in Sec. \ref{SecIIA}, which may differ from the formula used in Refs. \onlinecite{Morell_2017,Stauber_2018}, where $\text{CD}$ is simply proportional to the real part of $\sigma_{xy}$.

\section{Conclusion}\label{SecIV}
The field of engineering two-dimensional quantum materials for electronic and optoelectronic applications has been the subject of extensive research. In this work, we demonstrate that the optical activity observed in the TBG system results from the spatial dispersion effects of a typical vdW heterostructure of two graphene layers lacking mirror and glide symmetries. Our analysis addresses two fundamental aspects of the field-matter interaction problem: the propagation of electromagnetic waves through a material layer and the material layer's response to an external field, both treated simultaneously using the macroscopic and microscopic descriptions, respectively.

Technically, we utilize an effective continuum model to demonstrate how electron states in each graphene layer can hybridize, forming electron states in the bilayer system that allow for the correlation of transverse motions between the two layers. In terms of electromagnetic wave propagation, we clarify how the coupling between the two layers can transform the magnitude and direction of the fields within the material layer. We provide a detailed solution to the electromagnetic wave propagation problem through a TBG sheet by considering each graphene layer as a conducting interface between two dielectric media. The two interfaces are not independent and conduct mutually influenced current densities. Our solution to the wave propagation problem clarifies how the roles of the conductivity components characterized by local and nonlocal spatial effects manifest themselves in the optical behavior of the material system.

On the basis of explicitly considering the spatial dispersion effects, our analysis naturally composes the conductivity tensor into the local part and the drag part without relying on symmetry analysis or assumptions about the phase shift between current operators in the two layers. We calculate all elements of these parts using the Kubo formula, and formulate and calculate the transfer matrix numerically. We demonstrate that the transverse drag transverse conductivity $\sigma_{xy}^{(drag)}$ plays a critical role in defining the chiral optical response of the TBG system. This conductivity arises from the chiral structure of the atomic lattice, which leads to hybridized states that lack mirror symmetry and support the correlation of transverse motions between the two graphene layers. We use the results of our calculations of the components of the optical conductivity tensor to calculate the transmission and circular dichroism spectra. We show that the overall transparency of twisted bilayer graphene (TBG) samples, on average, is generally 98\%, but it has the potential to absorb light up to 2\%. Significantly, we establish that the drag transverse conductivity is a decisive factor in determining the circular dichroism, thereby affirming that the optical activity of the TBG system is a manifestation of spatial effects.

Furthermore, we found that the DC conductivity of the TBG system exhibits a quantum conductivity value of $\propto e^2/h$ at the intrinsic Fermi energy, which is determined by the hybridized states in the bilayer system. This quantum value is obtained from a single-particle approximation. However, a rigorous discussion of the intrinsic transport properties of TBG configurations with magic twist angles should be based on a strong correlation picture of low-energy excited states as Dirac fermions. In summary, our study emphasizes the importance of considering the finite thickness of twisted bilayer graphene when analyzing its optical response and transport properties, owing to the significance of spatial dispersion effects. This theoretical approach can also be extended to other van der Waals material systems with multiple layers.

\section*{Acknowledgements}
One of the authors, S.T.H., acknowledges the financial support of Hanoi University of Civil Engineering (HUCE) for his work under grant number 28-2021/KHXD-TD.

\bibliographystyle{apsrev4-1}
\bibliography{bibliography}

\end{document}